\documentclass{aa}
\usepackage{epstopdf}
\usepackage{graphicx}
\usepackage{array}
\usepackage{lscape}                                
\usepackage{natbib}
\usepackage{longtable}
\usepackage{color}
\usepackage{mathrsfs} 
\usepackage{txfonts,textcomp}
\usepackage{tikz}
\usepackage{amssymb}
\usepackage{multirow}
\usepackage{pdflscape}
\newcommand{\mom}{$\langle v \rangle$}
\newcommand{\kms} {km\,s$^{-1}$}
\newcommand{\vsini} {$v$\,sin\,$i$}

\newcommand{\vmac} {$v_{\rm mac}$}
\newcommand{\Teff} {T$_{\rm eff}$}
\newcommand{\grav} {log\,{\em g}}
\newcommand{\fastwind} {{\sc fastwind}}

\begin{document}

  \title{Low-frequency photospheric 
and wind variability in the early-B supergiant HD\,2905 
  }

   \author{S.~Sim\'on-D\'{\i}az\inst{1,2}, C.~Aerts\inst{3,4,5}, M.A.~Urbaneja\inst{6},
   I.~Camacho\inst{1,2}, V.~Antoci\inst{7}, M.~Fredslund Andersen\inst{7}, F.~Grundahl\inst{7}, P.L.~Pall\'e\inst{1,2}
   }

\institute{
Instituto de Astrof\'isica de Canarias, E-38200 La Laguna, Tenerife, Spain              
\and
Departamento de Astrof\'isica, Universidad de La Laguna, E-38205 La Laguna,
Tenerife, Spain
\and
Instituut voor Sterrenkunde, KU Leuven, Celestijnenlaan 200D, B-3001 Leuven,
Belgium
\and
Department of Astrophysics, IMAPP, University of Nijmegen, P.O. Box 9010, 6500
GL Nijmegen, The Netherlands
\and
Kavli Institute for Theoretical Physics, University of California, Santa
Barbara, CA 93106, USA
\and
Institut für Astro- und Teilchenphysik, Universität Innsbruck, Technikerstr. 
25/8, A-6020 Innsbruck, Austria
\and
Stellar Astrophysics Centre (SAC), Department of Physics and Astronomy, Aarhus University, Ny Munkegade 120, DK-8000 Aarhus, Denmark
}

\offprints{ssimon@iac.es}

\date{Submitted/Accepted}

\titlerunning{Low-frequency photospheric and wind variability in HD\,2905}

\authorrunning{S. S.-D. et al.}

 
  \abstract
  {Despite the important advances in space asteroseismology during the last
    decade, the early phases of evolution of stars with masses above
    $\sim$15~M$_{\odot}$ (including the O stars and their evolved descendants,
    the B supergiants) have been only vaguely explored up to now.  This is due
    to the lack of adequate observations for a proper characterization of
    the complex spectroscopic and photometric variability occurring in these
    stars.}
  {Our goal is to detect, analyze and interpret variability in the early-B type
    supergiant HD\,2905 ($\kappa$~Cas, B1~Ia) using long-term, ground based, high
    resolution spectroscopy.}
  {We gather a total of 1141 high-resolution spectra covering some 2900 days with
    three different high-performance spectrographs attached to 1\,--\,2.6m
    telescopes at the Canary Islands observatories. We complement these
    observations with the $Hipparcos$ light curve, which includes 160 data
    points obtained during a time span of $\sim$1200 days. We investigate
    spectroscopic variability of up to 12 diagnostic lines by using the zero and
    first moments of the line profiles. We perform a frequency analysis of both
    the spectroscopic and photometric dataset using Scargle periodograms. We
    obtain single snapshot and time-dependent information about the stellar
    parameters and abundances by means of the {\fastwind} stellar atmosphere
    code.}
  {HD\,2905 is a spectroscopic variable with peak-to-peak amplitudes in the zero
    and first moments of the photospheric lines of up to 15\% and 30 \kms,
    respectively. The amplitude of the line-profile variability is correlated
    with the line formation depth in the photosphere and wind. All investigated
    lines present complex temporal behavior indicative of multi-periodic
    variability with timescales of a few days to several weeks. No short-period
    (hourly) variations are detected.  The Scargle periodograms of the $Hipparcos$
    light curve and the first moment of purely photospheric lines
    reveal a low-frequency amplitude excess and a clear dominant frequency at
    $\sim$0.37 d$^{-1}$. In the spectroscopy, several additional frequencies
    are present in the range 0.1\,--\,0.4 d$^{-1}$.  These may be associated 
    with heat-driven gravity modes, convectively-driven gravity waves, or 
    sub-surface convective motions.  Additional frequencies are detected below 
    0.1\,d$^{-1}$. In the particular case of H$\alpha$, these are produced by 
    rotational modulation of a non-spherically symmetric stellar wind.}
  {Combined long-term uninterrupted space photometry with high-precision 
   spectroscopy is the best strategy to unravel the complex low-frequency 
   photospheric and wind variability of B supergiants. 3D simulations
   of waves and of convective motions in the sub-surface layers can shed 
   light on a unique interpretation of the variability.}


   \maketitle

\section{Introduction}\label{intro}

Asteroseismology has witnessed an immense progress in the last decade thanks to
the considerable increase in the amount of available high-precision
uninterrupted space photometry (e.g., Chaplin \& Miglio 2013; Charpinet et
al. 2014; Aerts 2015; Hekker \& Christensen-Dalsgaard 2016, for reviews).

Within the first part of this asteroseismic (space) revolution -- mainly driven
by the MOST, CoRoT and {\em Kepler} missions --, massive O-type stars and their
evolved descendants, the B supergiants (Sgs), were only serendipitously
targeted. As a consequence, just a handful of these stars could be investigated
in detail using photometric data from these satellites \citep{Saio06,
  Degroote10, Aerts10, Blomme11, Briquet11, Mahy11, Moravveji12, Aerts13,
  Aerts17}. This has hampered the possibility to perform a proper global
assessment of the main seismic properties of the early phases of evolution of
stars with masses above $\sim$15~M$_{\odot}$.  The situation is expected to
improve soon, thanks to new observations obtained by the BRITE-Constellation
satellites and the K2 mission \citep[see already some examples in][]{Buy15,
  Buy17, Pablo17}. However, the definitive breakthrough in the O star and B
supergiant domain will (hopefully) only become a reality after the launching of
the TESS and PLATO space missions.

Meanwhile, important advances can be also achieved using ground-based
time-resolved spectroscopy \cite[e.g.,][]{Kraus15, Aerts17}. Indeed, these
complementary observations will become decisive for disentangling the intricate
-- and most likely interconnected -- photospheric and wind variability present
in the whole massive star domain \cite[see][for some illustrative
examples]{Ebbets82, Prinja96, Fullerton96, Kaufer96, Kaufer97, Rivinus97,
  Morel04, Markova05, Markova08, Martins15}. Such data are in any case needed to
perform a reliable seismic characterization of these extreme stellar objects.
For this reason, we started long-term spectroscopic monitoring of about 20
bright Galactic O stars and B Sgs using the high-resolution spectrographs
attached to the Nordic Optical telescope (NOT\,2.56m), the Mercator\,1.2m
telescope, and the Hertzsprung-SONG\,1m telescope, all of them operated in the
Canary Islands observatories\footnote{Some observations of this joint project
  have been also obtained with the T13\,2.0m Automatic Spectroscopic Telescope
  (AST) operated by Tennessee State University at the Fairborn Observatory and
  the second SONG node operated in Delingha (China).}.
First results of this collaboration were presented in \cite{Aerts17}, where we
focused on the detection, analysis and interpretation of the photospheric and
wind variability of the late-O supergiant HD~188209 as deduced from {\em Kepler\/}
space photometry and long-term high-resolution spectroscopy. Here, we
investigate variability in the early-B type Supergiant HD\,2905 ($\kappa$~Cas,
B1~Ia) mainly using ground-based high-resolution spectroscopy obtained during
$\sim$2900 days.

The paper is structured as follows. After describing in Sect.~\ref{section2} the
main characteristics of the compiled spectroscopic dataset, we summarize in
Sect.~\ref{section3} the mean stellar parameters and abundances of HD\,2905. This
work mostly concentrates on the characterization of the spectroscopic
variability present in this early-B Supergiant; however, we also analyze the
available {\em Hipparcos\/} light curve to investigate its photometric
variability. The results obtained from both independent analyses
are presented in Sect.~\ref{section4}, while we provide a discussion 
on the physical interpretation of the observed variability in Sect.~\ref{section5}.

\begin{table*}
  \caption{Summary of the spectroscopic and {\em Hipparcos} observations used in this work, where the HJD is listed with respect to
    2,400,000. The listed Nyquist frequency is the inverse of twice the shortest time interval between two consecutive measurements. The resolving power associated with each instrumental configuration is $R$\,=\,46000 (FIES), 85000 (HERMES), and 77000 (SONG).}
	\label{obslog} 
	\centering 
	\begin{tabular}{rrrrrrr}
\hline\hline
\noalign{\smallskip}
Instrument     & HJD start & HJD end & N  & $\Delta$T & Rayleigh
          & Nyquist \\ 
 & [d] & [d] &  & [d] & [d$^{-1}$] & [d$^{-1}$] \\
          \hline
		\noalign{\smallskip}
FIES-1    & 54776.37 & 54779.44 &  30 &    3.07  & 0.16281 &  654 \\
FIES-2    & 55145.30 & 55148.60 &  28 &    3.30  & 0.15162 &  115 \\
FIES-3    & 57592.52 & 57595.74 &  17 &    3.22  & 0.15532 &   66 \\
		\hline
		\noalign{\smallskip}
HERMES-1  & 56590.34 & 56596.61 &  25 &    6.27 & 0.07972 &   21 \\
HERMES-2  & 56640.31 & 56649.51 &  56 &    9.20 & 0.05436 &   17 \\
HERMES-3  & 56909.36 & 56911.65 &  33 &    2.29 & 0.21804 &  125 \\
HERMES-4  & 57576.56 & 57580.73 &  16 &    4.17 & 0.11992 &   20 \\
HERMES-5  & 57658.40 & 57663.64 &  26 &    5.25 & 0.09530 &   11 \\
HERMES-6  & 57679.30 & 57684.54 &  27 &    5.24 & 0.09545 &   12 \\
		\hline
		\noalign{\smallskip}
SONG-1    & 57263.39 & 57270.76 & 447 &    7.37 & 0.06782 &   89 \\
SONG-2    & 57586.64 & 57606.73 &  45 &   20.09 & 0.02489 &   82 \\
SONG-3    & 57607.52 & 57617.74 & 389 &   10.22 & 0.04892 &   89 \\
		\hline
		\hline
		\noalign{\smallskip}
FIES      & 54776.37 & 57595.74 &  77 & 2819.36 & 0.00018 &  654 \\
HERMES    & 56590.34 & 57684.54 & 183 & 1094.20 & 0.00046 &  125 \\
SONG      & 57263.39 & 57617.74 & 881 &  354.36 & 0.00141 &   89 \\
		\hline
		\hline
		\noalign{\smallskip}
{\em Hipparcos} & 47868.79 & 49042.53 & 160 & 1173.74 & 0.00085 & 35\\
\hline
	\end{tabular}
\end{table*}

\section{Observations}\label{section2}

The bulk of the observations considered in this paper has been gathered in the
framework of the IACOB project \citep{Simon11, Simon15}. Our monitoring of
HD\,2905 started with a couple of 4-night runs with the FIES instrument
\citep{Tel14} attached to the 2.56-m Nordic Optical Telescope (NOT, Observatorio
del Roque de los Muchachos, La Palma, Spain). These spectra were used in
\cite{Simon10} to investigate the possible connection between macroturbulent
broadening and line-profile variations in OB supergiants. HD\,2905 was one of the
targets in the sample for which a large spectroscopic variability was detected,
with a peak-to-peak amplitude of the first moment\footnote{For a definition, see
  Aerts et al. (2010, Chapter 4).} of the \ion{Si}{iii}$\lambda$4552 line
somewhat larger than 10~\kms. We hence decided to carry on with the
spectroscopic monitoring of this star, performing several additional dedicated
runs with the HERMES spectrograph \citep{Ras11} attached to the 1.2-m Mercator
telescope (Observatorio del Roque de los Muchachos) and the \'echelle
spectrograph attached to the prototype 1-m Hertzsprung-SONG node \citep{Gru07,
  Gru14} operational at Observatorio del Teide (Tenerife, Spain).

We paid specific attention to assemble an overall spectroscopic data set that
contains stable high-precision spectroscopy with a long time base (FIES and
HERMES data), as well as high cadence (SONG) spectroscopy. These two aspects are
essential if one wants to achieve an overall view of long-term low-amplitude
supergiant variability \citep[see, e.g.,][]{Aerts17}.

\begin{figure}[!t]
\centering
\includegraphics[width=0.47\textwidth,angle=0]{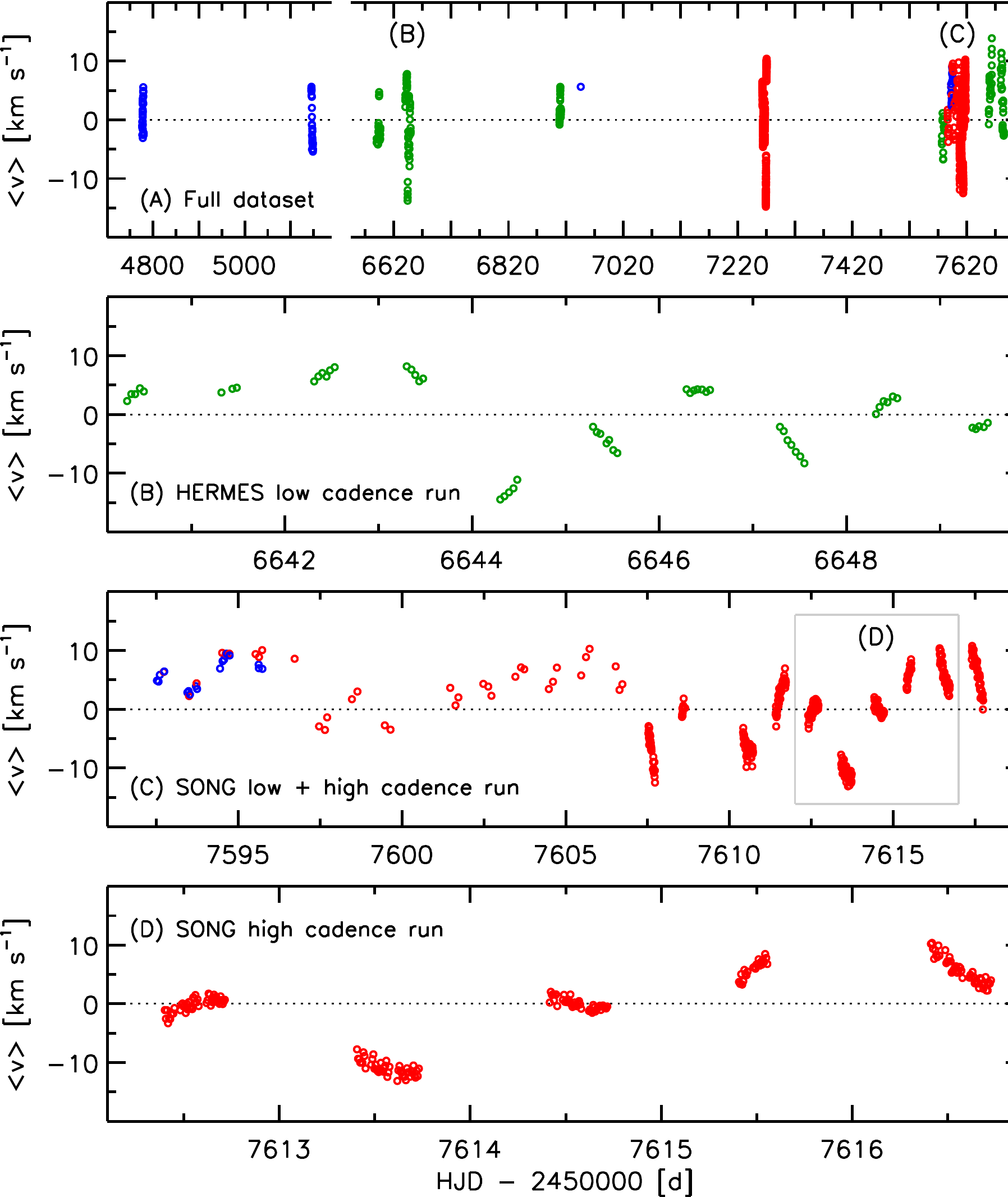}
\caption{Summary of spectroscopic observations compiled for HD\,2905 with the FIES (blue), HERMES (green) and SONG (red) spectrographs. The various panels show the variability of the first normalized moment (centroid in velocity) of the \ion{Si}{iii}$\lambda$4552 line.}
\label{fig1}
\end{figure}

A summary of the spectroscopic observations used in this work can be found in
Table~\ref{obslog}. We first provide information about the separated runs and
then the global characteristics of the data~set associated with each
instrument. We also present a visual overview of the compiled observations in
Fig.~\ref{fig1}. We gathered a total of 1141 high~resolution (R) high
signal-to-noise (S/N) spectra during some 2900~d separated in various observing
runs ranging from 3 days to about 1 month. Most of these observing runs
correspond to HERMES and FIES spectra obtained in low cadence mode
(5\,--\,15~spectra per night with a separation of 1\,--\,2~h during
3\,--10~nights -- see, e.g., panel B in Fig.~\ref{fig1}). We also benefited from
the much more versatile operational mode of the SONG telescope to perform 2 runs
in high cadence mode (8 and 11 consecutive nights, respectively, with a cadence
of $\sim$8~min and almost full night coverage -- see, e.g., panels C and D in
Fig.~\ref{fig1}). The second SONG high cadence run was allocated just after a
longer SONG low cadence mode (21 nights, 2\,--\,3 spectra per night). In all
cases, HD\,2905 was observed when the star had a visibility of at least 3~h and a
maximum of $\sim$10~h.

The observing strategy was different between the SONG and FIES+HERMES
campaigns. In the former, the exposure time was fixed to 480\,s. This resulted
in a S/N ranging from 75 to 200 depending on the altitude of the star and the
sky conditions. In the latter, we use the exposure tool available at the
telescope to end up with a spectrum with S/N\,$\sim$\,200\,--\,300. This implied
a typical exposure time of $\sim$150\,s (FIES) and $\sim$200~s (HERMES) and a
range in this quantity of 100\,--\,500\,s depending on the sky conditions
when the star was observed.

Our on-ground spectroscopic observations were complemented with the
publicly available {\it Hipparcos\/} light curve of HD\,2905 containing 160 datapoints
assembled during $\sim$3.2 years (see details in the last row of
Table~\ref{obslog}).

\section{Mean stellar parameters and abundances}\label{section3}

\begin{table*}
  \caption{Stellar+wind parameters and abundances of HD\,2905. The 2nd and 3rd 
    columns
    correspond to the outcome of our {\sc FASTWIND} 
    analysis of the mean of 56 spectra obtained during the 
    HERMES-2 run (central values and associated uncertainties),
    while the 4th, 5th, 6th columns list the results from  
    the individual spectra 
    (mean value, standard deviation, minimum and maximum values 
    of individual estimates). We also include the set of parameters and 
    abundances determined by \cite{Crowther06} and \cite{Searle08} 
    for comparison purposes.} 
\label{parameters_log} 
\centering 
\begin{tabular}{l rl rll crrl}
\hline\hline
\noalign{\smallskip}
                        & \multicolumn{5}{c}{This work} & & \multicolumn{2}{c}{Literature} & \\
       \cline{2-6} \cline{8-9} 
\noalign{\smallskip}
                        & \multicolumn{2}{c}{Averaged} & \multicolumn{3}{c}{Multi epoch} & & C06 & S08 & Unit\\
\hline\hline
\noalign{\smallskip}
 \vsini                  &  58    & $\pm$\,5      & 59   & $\pm$\,6      & [48,70]            &  & 91$^{\rm (a)}$  & 91$^{\rm (a)}$ & \kms \\
 \vmac                   &  82    & $\pm$\,6      & 81   & $\pm$\,8      & [68,95]            &  & ---    & ---    & \kms \\
\noalign{\smallskip}
 \Teff                   & 24.6   & $\pm$\,0.3    & 24.5 & $\pm$\,0.3    & [23.7, 25.1]     &  & 21.5   & 23.5   & kK \\ 
 \grav                   & 2.79   & $\pm$\,0.05       & 2.79 & $\pm$\,0.05   & [2.68, 2.92]     &  & 2.60   & 2.75   & dex \\ 
 N(He)/N(H)              & 0.11   & $\pm$\,0.02   & 0.10 & $\pm$\,0.02   & [0.08, 0.14]     &  & ---    & ---    &  \\ 
 log~$Q$                 & -12.73 & $\pm$\,0.05   & -12.74 & $\pm$\,0.05 & [-12.83, -12.66] &  & -12.7$^{\rm (b)}$ & -12.4$^{\rm (b)}$ &  \\ 
 $\beta$                 & 1.56   & $\pm$\,0.10   & 1.55 & $\pm$\,0.20   & [1.24, 1.96]     &  & 2.0    & 1.5    &  \\ 
 $\xi_{\rm t}$           & 19.0   & $\pm$\,1.0    & 19.9 & $\pm$\,1.2    &  [18.4, 22.5]    &  & 20     & 20     & \kms\ \\ 
\noalign{\smallskip}
 log(Si/H)+12            & 7.84   & $\pm$\,0.10   & 7.79 & $\pm$\,0.14   & [7.50, 7.96]     &  & ---    & ---    & dex \\ 
 log(N/H)+12             & 7.69   & $\pm$\,0.08   & 7.67 & $\pm$\,0.04   & [7.59, 7.73]     &  & 8.00   & 8.16   & dex \\ 
 log(O/H)+12             & 8.89   & $\pm$\,0.05   & 8.87 & $\pm$\,0.04   & [8.77, 8.93]     &  & 8.75   & 8.80   & dex \\ 
 log(Mg/H)+12            & 7.43   & $\pm$\,0.09   & 7.42 & $\pm$\,0.04   & [7.34, 7.48]     &  & ---    & ---    & dex \\ 
\noalign{\smallskip}
 log$L/L_{\odot}$        &  5.69  & $\pm$\,0.12$^{\rm (c)}$ &      & ---           &             &  & 5.52   & 5.48   & \\
 $\dot{M}$ &  1.7   & $\pm$\,0.6$^{\rm (c)}$  &      & ---
                        &             &  & 2.0    &  2.5   & $10^{-6} M_{\odot}$~yr$^{\rm -1}$ \\ 
 $R$                     &  39    & $\pm$\,5$^{\rm (c)}$    &       & ---           &             &  & 41.4   & 33.0   & $R_{\odot}$ \\
\hline
\noalign{\smallskip}
 \multicolumn{10}{l}{\small $^{\rm (a)}$ Adopted from \cite{Howarth97}. This value assumes that all line-broadening is due to rotation.} \\
 \multicolumn{10}{l}{\small $^{\rm (b)}$ Computed using the values of $\dot{M}$ and $R$ in the Table, plus $v_{\infty}$\,=\,1105~\kms\ \citep{Prinja90}.} \\
 \multicolumn{10}{l}{\small $^{\rm (c)}$ Computed using the values of \Teff, \grav\ and log$Q$ in the Table, plus $v_{\infty}$\,=\,1105~\kms\ and $M_{\rm v}$\,=\,-7.1 (see text).} \\
\end{tabular}
\end{table*}

Given the visual magnitude of HD\,2905 (V\,=\,4.16 mag), 
this star occurs in most of the studies investigating the physical properties of
relative large samples of Galactic B supergiants \citep[see,
e.g.,][]{McErlean99, Kudritzki99, Crowther06, Searle08}. In this paper, we
performed our own determination of the basic stellar and wind parameters, as
well as abundances with a twofold objective. We wanted to (1) revisit the
outcome from the quantitative spectroscopic analysis of this early-B supergiant
using observations of much better quality than previous studies and (2)
investigate whether the detected spectroscopic variability is also reflected in
the quantities resulting from the analysis of individual spectra.

To this aim, we considered the 56 spectra corresponding to the HERMES-2 run (see
Table~\ref{obslog} and the second row in Fig.~\ref{fig1}) and one additional
spectrum with a much higher S/N obtained from adding all these spectra. For the
estimation of the stellar+wind parameters and abundances we relied on the
non-LTE atmosphere/spectrum synthesis code {\sc fastwind} \citep{Santolaya97, 
Puls05, Rivero11} and a quantitative spectroscopic analysis strategy.
Prior to the {\sc fastwind} analysis, we used the
{\sc iacob-broad} tool \citep{Simon14} to perform a line-broadening analysis of
the \ion{Si}{iii}~$\lambda$4552 line in each of the 56 individual and the
averaged spectra in order to obtain estimates for the projected rotational
velocity (\vsini) and the amount of macroturbulent broadening (\vmac).

In addition to \vsini\ and \vmac, the other quantities directly obtained from
the spectroscopic analysis are the effective temperature (\Teff), the stellar
gravity (\grav), the wind-strength parameter ($Q$) and the exponent of the
wind-velocity law ($\beta$), plus the microturbulence ($\xi_{\rm t}$) and the
absolute helium, silicon, oxygen and magnesium surface abundances. To obtain
estimates of all these quantities, we 
basically followed the same fitting
strategy (including the set of diagnostic lines) as described in \cite{Urbaneja17}
and utilized a similar grid of {\fastwind} models (but for solar metallicity). 
Despite some degree of
variability in the derived values of the two line-broadening parameters (see
Table~\ref{parameters_log}), we decided to fix these two quantities to
\vsini\,=\,58~\kms\ and \vmac\,=\,82~\kms\ in the {\sc fastwind} analysis of the
complete dataset.

The spectroscopic parameters were then combined with information about the
terminal velocity ($v_{\infty}$) of the stellar wind and the absolute magnitude
in the V-filter ($M_{\rm v}$) to obtain estimates for the stellar radius ($R$),
mass-loss rate ($\dot{M}$) and luminosity ($L$). In particular, we adopted the
robust UV determination of the wind velocity inferred by \cite{Prinja90} and
\cite{Howarth97} from the analysis of the IUE high resolution spectrum of this
star ($v_{\infty}$\,=\,1105~\kms) and the nominal distance modulus to Cas\,OB14
(to which HD\,2905 belongs) proposed by \cite{Humphreys78}. This distance
($m$-$M$\,=\,10.2) was then combined with our own estimation\footnote{We assumed
V\,=\,4.16 mag, B\,=\,4.30 mag \citep{Ducati02}, $R_{\rm v}$\,=\,3.1, and the
intrinsic color excess (B-V)$_{\rm 0}$\,=\,-0.21 mag obtained from {\sc
fastwind} model with stellar parameters indicated in the first column of
Table~\ref{parameters_log}.} of the extinction in the V filter ($A_{\rm v}$)
to end up in $M_{\rm v}$\,=\,-7.1 mag.

\begin{figure}[!t]
\centering
\includegraphics[width=0.48\textwidth,angle=0]{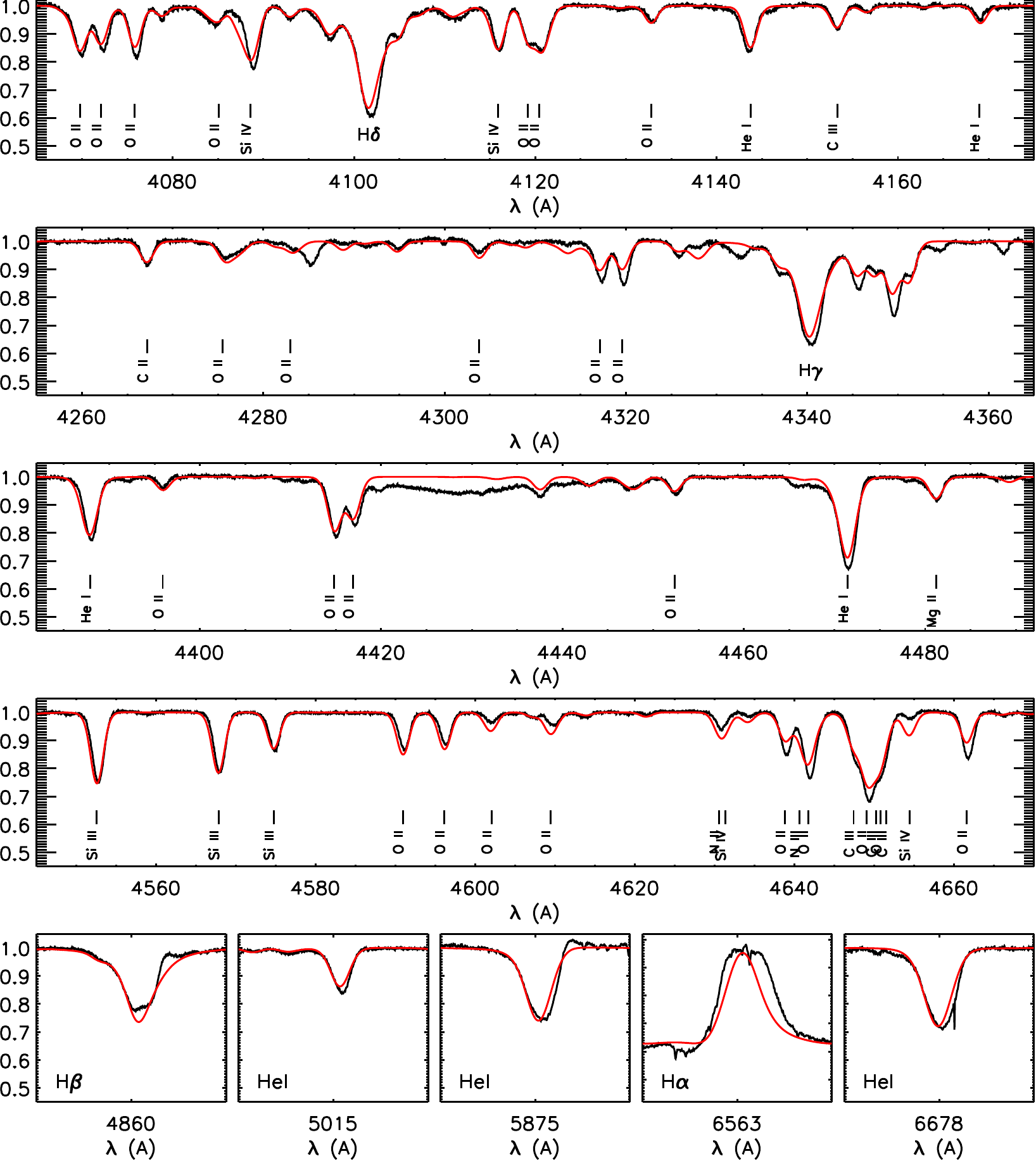}
\caption{Portions of the spectrum of HD\,2905 showing the quality of the fit of the {\sc fastwind} synthetic spectrum (red) to one of analyzed spectra (black). Main diagnostic lines providing information about stellar parameters, abundances and/or spectroscopic variability also indicated.}
\label{fig2}
\end{figure}

Table~\ref{parameters_log} summarizes the outcome of our
quantitative spectroscopic analyses applied to the averaged and individual
spectra, respectively. We indicate in the 2nd and 3rd column the central
values and associated uncertainties resulting from the analysis of the averaged
spectrum. The 4th, 5th, and 6th columns quote the mean and standard deviation
(as well as the minimum and maximum values) obtained from the individual 56
spectra of the HERMES-2 run.  The latter, once compared with the intrinsic
uncertainties of the analysis strategy, provides a rough indication of potential
variability in the derived parameters. The above mentioned information is
complemented with Fig.~\ref{fig2}, where we show the quality of the final fit of
the corresponding {\sc fastwind} synthetic spectrum to one of the analyzed
spectra. In addition, and for comparison purposes, we also include the set of
stellar parameters and abundances obtained by \cite{Crowther06} and
\cite{Searle08}, based on the analysis of an intermediate resolution spectrum
covering a more limited wavelength range \citep{Lennon92} with the stellar
atmosphere code {\sc cmfgen} \citep{Hillier98, Hillier03}.
The agreement between the results from the analysis of the combined spectrum
(central values and formal uncertainties) and the time-series (mean values and
standard deviation) is remarkably good. However, there is a non-negligible scatter
in the quantities resulting from the analysis of the individual spectra that may be
pointing towards a real variability in the derived paramters and abundances.

\section{Detection and characterization of the stellar variability}
\label{section4}

\subsection{Photometric variability}\label{hipparcos}

In contrast to the case of HD\,188209 discussed in \citet{Aerts17}, we do not
have high-cadence space photometry at our disposal. We thus considered the {\em
  Hipparcos\/} light curve as the best alternative to investigate the photometric
variability. Despite the fact that this only allows to search for variability at
mmag level, the {\em Hipparcos\/} data proved to be a suitable resource to
find gravity-mode pulsations of such amplitude in evolved hot supergiants
\citep[e.g.,][]{Lefever07}.

\begin{figure}[!t]
\centering
\includegraphics[width=0.32\textwidth,angle=90]{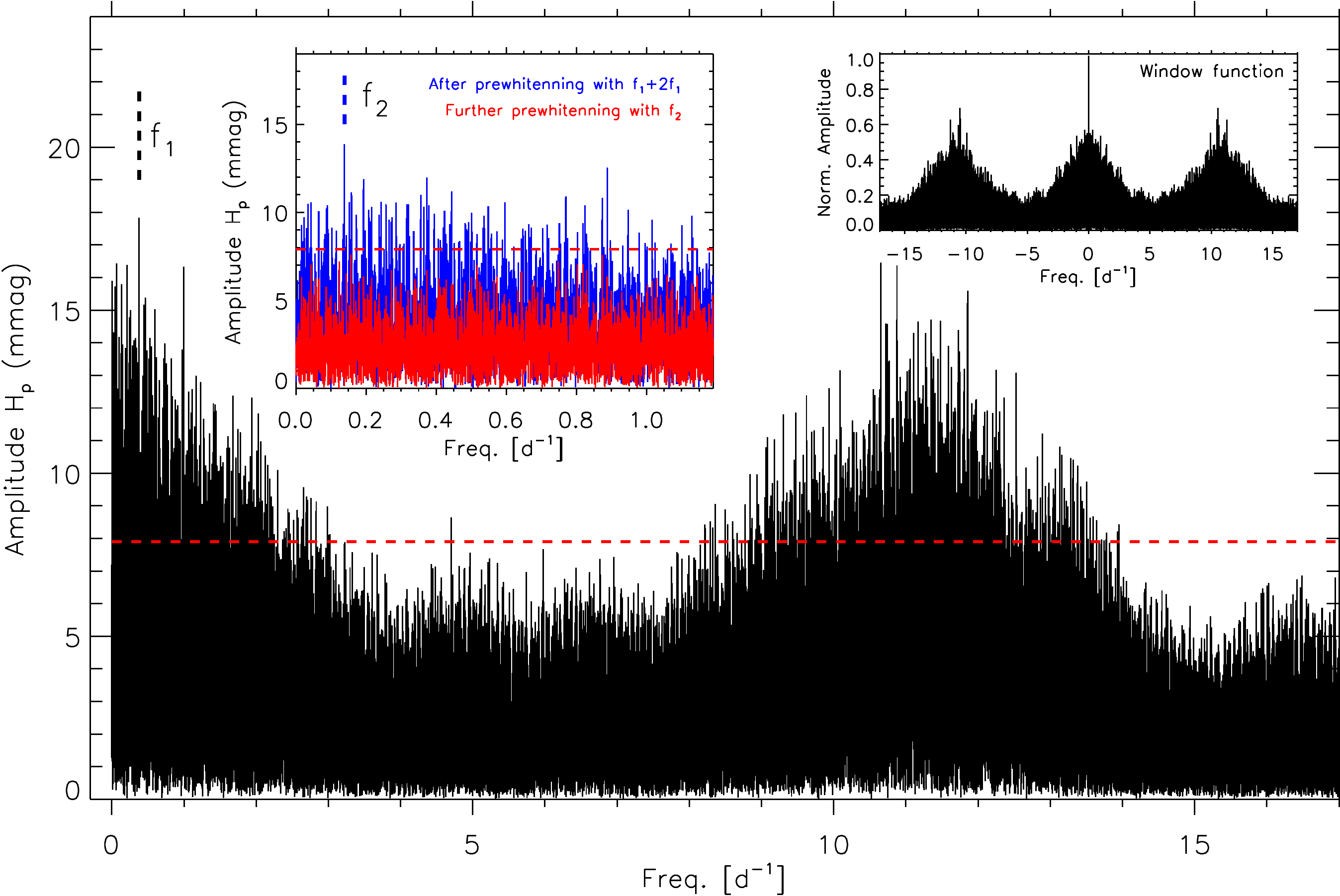}
\caption{Amplitude spectrum of the {\em Hipparcos\/} light curve of
  HD\,2905, where the dominant frequency peak ($f_1$) is also indicated. The right inset represents the spectral window. The left inset is a zoom of the amplitude spectrum showing two subsequent levels of prewhitening and the only additional reliable frequency peak ($f_2$) standing out above four times the average noise level, indicated by the horizontal red dashed line (see text for details).}
\label{Scargle-Hp}
\end{figure}

The morphology of the Scargle periodogram (Fig.~\ref{Scargle-Hp}) is quite
different from the ones of $\beta\,$Cep pressure-mode pulsators
\citep[e.g.,][]{Aerts00} but does bear some resemblance to those of slowly
pulsating B stars, for which the dominant coherent gravity-mode pulsations can
be retrieved from {\em Hipparcos\/} photometry \citep{DeCat02}. Unlike for those
stars, however, the {\em Hipparcos\/} periodogram of HD\,2905 shows a clear
amplitude excess at frequencies below 4\,d$^{-1}$ from which a few significant
frequencies stand out. We note that the second amplitude bump located between 8
and 14\,d$^{-1}$ is a consequence of the limited sampling of the {\em
  Hipparcos\/} data, which produces a spectral window as in the inset of
Fig.~\ref{Scargle-Hp}. The periodogram resembles the one of HD~188209
\citep[Fig. 4 in][]{Aerts17}, where we also showed how the photometric amplitude
spectrum of that late-O supergiant improves when relying on {\em Kepler\/}
scattered-light photometry with a similar time-span as the {\em Hipparcos\/}
dataset but with a much higher cadence (the number of data points is $\sim$200
times larger).

\begin{figure}[!t]
\centering
\includegraphics[width=0.24\textwidth,angle=-90]{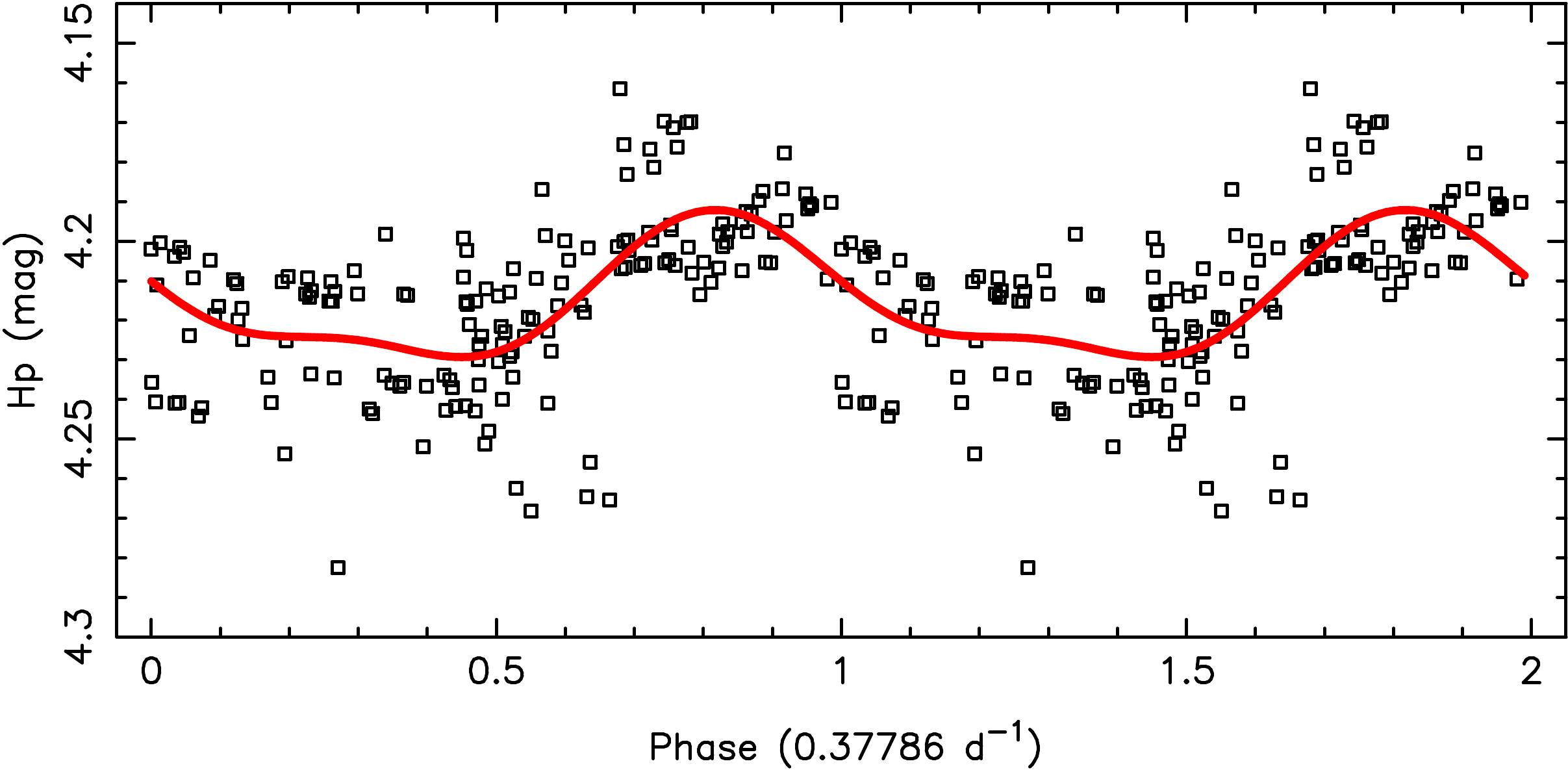}
\caption{{\em Hipparcos\/} light curve folded according to the dominant frequency. The
  red line is the fit for $f_1$ and its first harmonic.}
\label{Phase-Hp}
\end{figure}

On top of the general amplitude bump excess at low frequencies found from the
{\em Hipparcos\/} data, we find a dominant isolated frequency peak at
$f_1$\,=\,0.37786$\pm$0.00001~d$^{-1}$, with an amplitude of 18$\pm$2\,mmag (see
dashed line in the main panel of Fig.~\ref{Scargle-Hp}). A fit in the time domain
shows its first harmonic 2$f_1$ to have a significant amplitude as well, of
value 9$\pm$2\,mmag (this peak is buried within a forest of additional peaks in
the black curve in Fig.\,\ref{Scargle-Hp}). After prewhitening a harmonic fit 
with $f_1$ and $2f_1$, shown in Fig.\,\ref{Phase-Hp} as the red line,
we find a second frequency $f_2$\,=\,0.13813$\pm$0.00001~d$^{-1}$ in the
residuals
(indicted by
the blue dotted line in the left inset of Fig.\,\ref{Scargle-Hp}), with an amplitude of
14$\pm$2\,mmag.  Further prewhitening leads to
$f_3^\star$\,=\,11.27175$\pm$0.00002~d$^{-1}$ (located outside the frequency range shown in
the left inset of Fig.~\ref{Scargle-Hp}), with an amplitude of 8$\pm$2\,mmag. After prewhitening
this third frequency, we no longer find significant frequencies adopting the
common criterion of accepting a frequency if its amplitude is above four
times the average noise level in an oversampled periodogram \citep{Breger93}.
Even though it fulfils the noise criterion, the reliability of $f_3^\star$ is
not at similar level as the other two detected frequencies, especially since it
is located in a region of the Scargle periodogram affected by the
aliasing in the {\em Hipparcos\/} data.
The harmonic fit to the {\em Hipparcos\/} photometry with $f_1$ and $f_2$ explains a fraction of 74\% of
the variability in that light curve.

\subsection{Spectroscopic variability}\label{spec_var}

Subsequently, we turned to the time-resolved spectroscopy. 
Compared to the study of HD\,188209 performed in \cite{Aerts17}, more spectral
lines were available for the investigation of spectroscopic variability, thanks
to the somewhat lower effective temperature of HD\,2905. In the HERMES and FIES
spectra we had access to 13 non-blended -- and strong enough -- spectral lines,
among which \ion{Si}{iii}, \ion{Si}{iv}, \ion{O}{ii}, \ion{He}{i}, H$\beta$, and
H$\alpha$. In the SONG spectra, some of these lines are missing, mainly due to
the more limited wavelength coverage in the blue. In particular, the SONG
spectra lack the \ion{Si}{iv}~$\lambda$4116 line, an important diagnostic line
to establish the effective temperature of the star. Additional interest in this
line comes from the fact that, in early-B supergiants, it is formed in deeper
regions of the stellar photosphere compared to \ion{Si}{iii}, \ion{O}{ii}
and \ion{He}{i} lines.

\begin{figure}[!t]
\centering
\includegraphics[width=0.33\textwidth,angle=90]{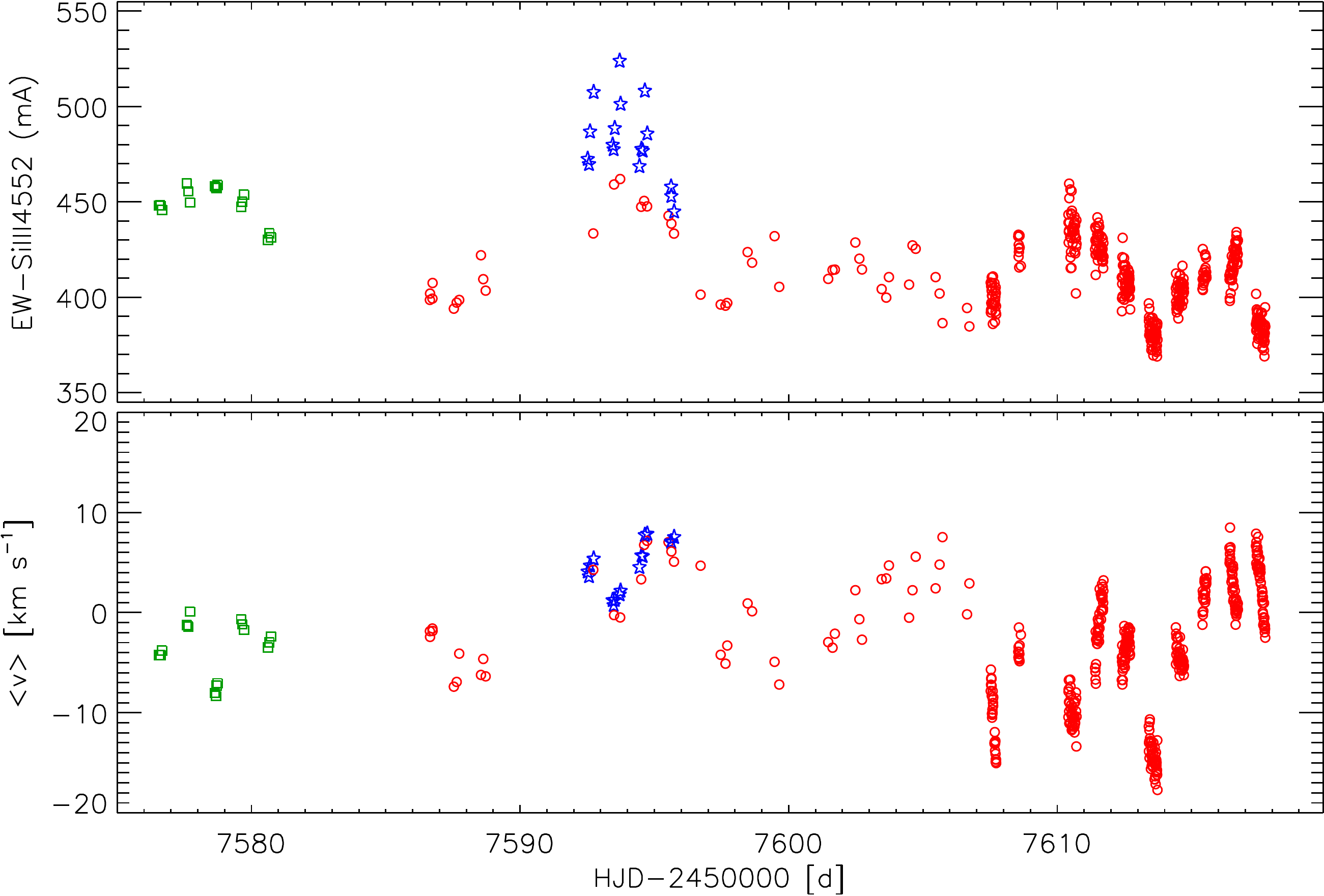}
\caption{The equivalent width (EW, top) and centroid of the line (\mom, bottom) of the \ion{Si}{iii}~$\lambda$4552 line. Blue: FIES, red: SONG, green: HERMES.}
\label{ew-vr}
\end{figure}

Following the strategy described in \cite{Aerts17}, we computed the equivalent
width (EW, moment of order zero expressed in \AA) and the centroid (first
moment, denoted as $\langle v \rangle$ expressed in \kms) for each of the lines
and for each of the three spectroscopic data sets listed in
Table\,\ref{obslog}. By way of example, we show the EW and $\langle v\rangle$ of
the \ion{Si}{iii}~$\lambda$4552\AA\ line in Fig.~\ref{ew-vr}. This line is the
deepest one of a triplet and was shown to be an optimal line-profile diagnostic
for low-amplitude variability of stars with similar temperature than HD\,2905
\citep[e.g.][]{Aerts03}. Despite careful normalization of all the spectra, it
can be seen in the top panel of Fig.~\ref{ew-vr} that there are non negligible
differences in the EW resulting from the three different spectrographs. This was
already stressed in \citet{Aerts17} and implies that simply merging data sets is
not optimal when hunting for low-amplitude variability. Thanks to the definition
of the line moments adopted in \citet{Aerts92}, differences in EW are
compensated for in \mom, as can be seen in the bottom panel of
Fig.~\ref{ew-vr}. Nevertheless, small differences due to the different
instruments and atmospheric conditions cannot be excluded, so we continued with
the three separate data sets as well as with their merged versions.

\subsubsection{Variability of zero and first moments
  of the line profiles}\label{moments-t}

A detailed look at Figs\,\ref{fig1} and \ref{ew-vr} reveals complex
temporal behavior of the \ion{Si}{iii}~$\lambda$4552 line profile of 
HD\,2905. The star is a clear spectroscopic variable with peak-to-peak amplitudes
in equivalent width and the first moment of this line up to 15\% and 30~\kms,
respectively.  We also conclude that we are dealing with 
multi-periodic variability, with timescales of a few to several days, in
qualitative agreement with $f_1$ and $f_2$ found in the {\em Hipparcos\/}
photometry. Visual inspection of the SONG high-cadence observations (e.g.,
bottom panel in Fig.~\ref{fig1}) does not reveal short-period variations.
This makes us conclude that any frequency peak detected in the {\em
  Hipparcos\/} periodogram above 2 d$^{-1}$ is likely spurious and a consequence
of the relatively poor time sampling of those photometric data.

\begin{figure}[!t]
\centering
\includegraphics[width=0.47\textwidth]{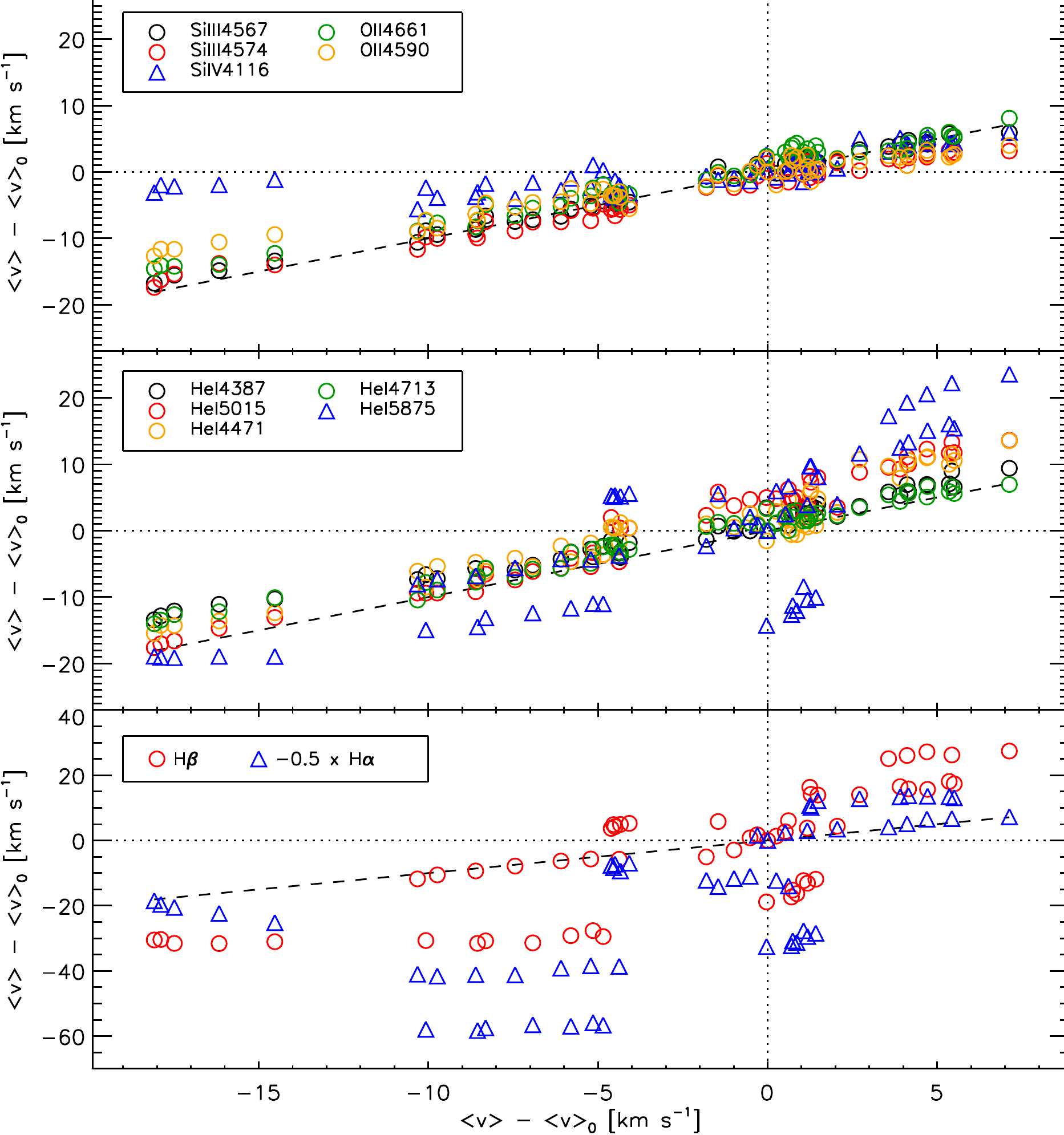}
\caption{Correlations in the temporal variability of the first moment (centroid) of 
13 diagnostic lines in HD\,2905. The \ion{Si}{iii}~$\lambda$4552 line is used as baseline (x-axis) and all 
measurements are referred to the value of \mom\ (per individual line) corresponding to the first spectrum in 
the HERMES-2 dataset (see also Table~\ref{obslog} and Fig.~\ref{vandew_ex}).}
\label{line2line_var}
\end{figure}

\begin{figure}[!t]
\centering
\includegraphics[width=0.47\textwidth]{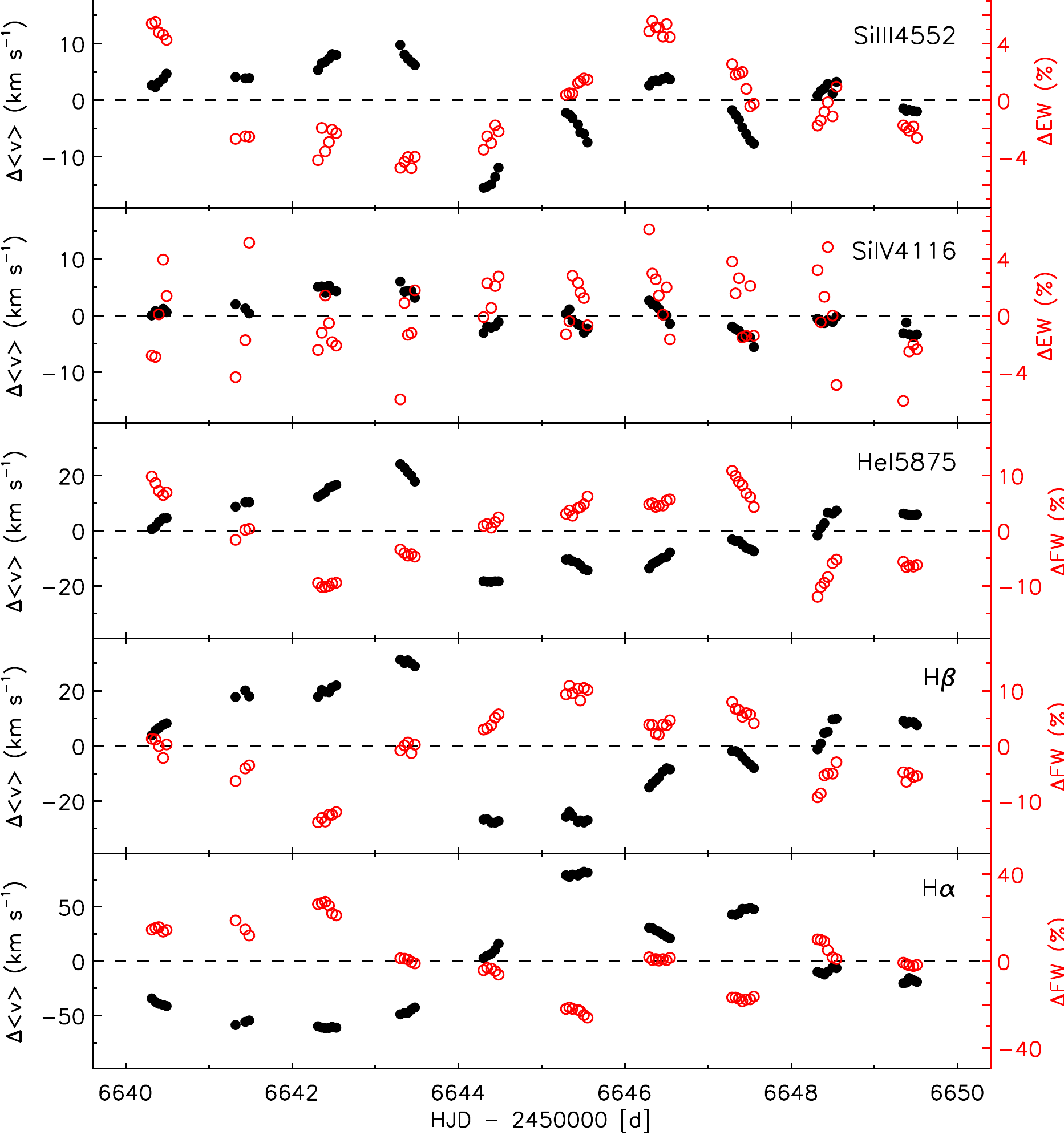}
\caption{Illustrative example of temporal variability of the zero (EW, red open circles) and first moment (\mom, black filled circles) of five representative  diagnostic lines in HD\,2905. The time baseline correspond to the 10 nights of observations obtained in the HERMES-2 run (see also Table~\ref{obslog} and Fig.~\ref{line2line_var}).}
\label{vandew_ex}
\end{figure}

In the following, we concentrate on data obtained during the HERMES-2 run (10
nights, see second panel in Fig.~\ref{fig1}) to illustrate the main conclusions
that can be extracted from the qualitative investigation of the temporal
behavior of the EW and centroid of the whole set of considered diagnostic
lines. As indicated above, we rely on this subset of spectra due to the
availability of the \ion{Si}{iv}~$\lambda$4116 line. Figures~\ref{line2line_var}
and \ref{vandew_ex} show the detected variability from two different
perspectives. On the one hand, Fig.~\ref{line2line_var} serves to illustrate
whether there exist correlations between temporal variations in the first
moment of the various lines. On the other hand, Fig.~\ref{vandew_ex} allows a
direct comparison of variability in EW and \mom\ as a function of time for a
subset of lines. For better visualization, we divide the results presented in
Fig.~\ref{line2line_var} in three panels including, from top to bottom, metallic
lines, \ion{He}{i} lines, and H$_{\alpha}$ and H$_{\beta}$. We use
\ion{Si}{iii}~$\lambda$4552 as baseline and refer all obtained values of \mom\
per line to the one corresponding to the first spectrum in the HERMES-2
time-series. This way we eliminate potential shifts in velocity between the
different lines due to atomic data (central wavelengths). Inspection of
Fig.~\ref{line2line_var} leads us to conclude that there is a fair correlation
between variability of \mom\ in all those lines which are mainly formed in the
same regions of the photosphere, i.e.\ all \ion{Si}{iii} and \ion{O}{ii} lines,
plus all \ion{He}{i} lines except for \ion{He}{i}~$\lambda$5875.  The latter,
similarly to H$_{\alpha}$ and H$_{\beta}$, seems to behave in a different way
and present a much larger variability. Last, there is a clear positive
correlation between the temporal variation of the centroid of the
\ion{Si}{iii}~$\lambda$4552 and \ion{Si}{iv}~$\lambda$4116 lines, but the
amplitude of variability is much smaller in the latter.

\begin{figure*}[!t]
\centering
\includegraphics[width=0.37\textwidth,angle=90]{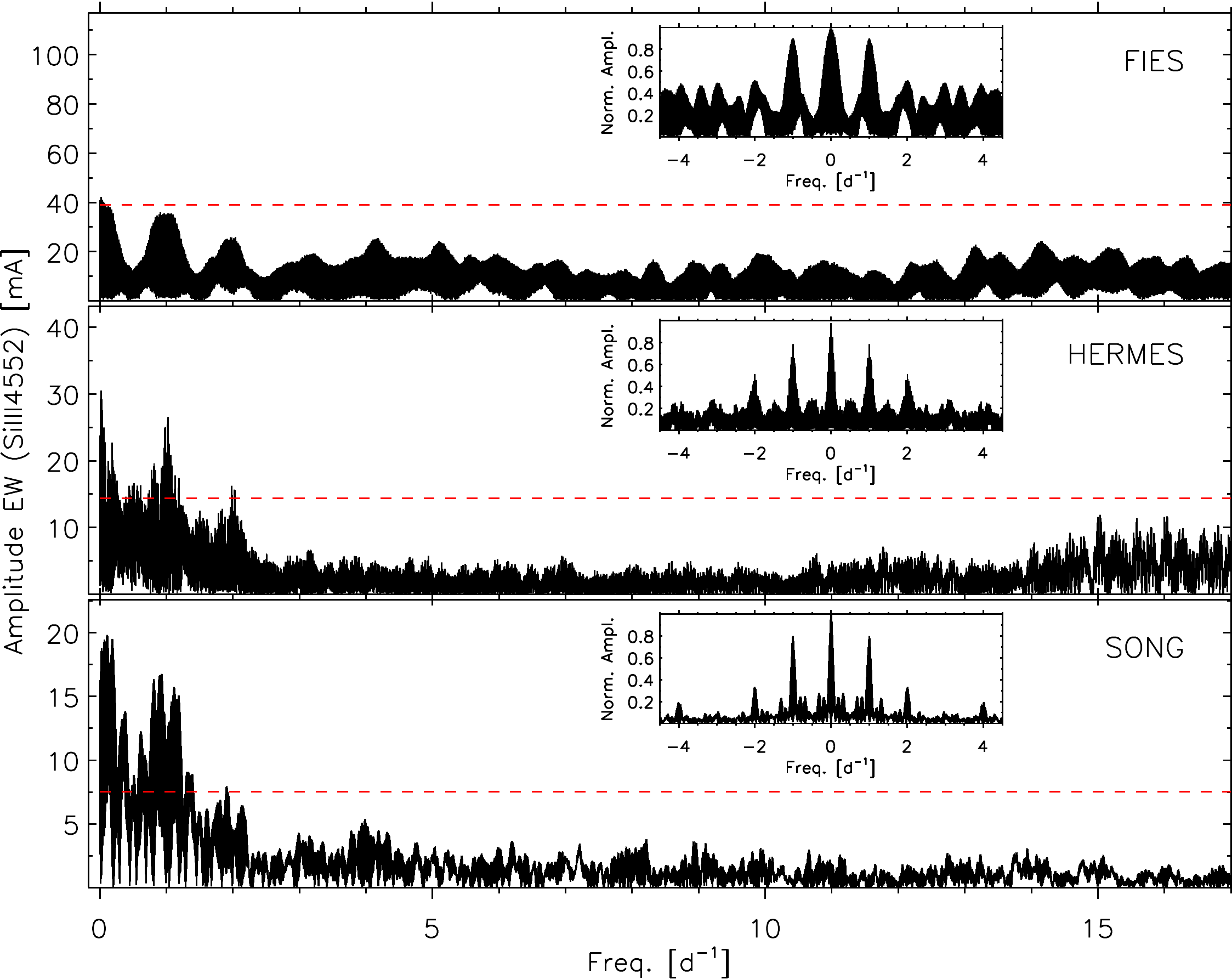}
\hspace*{0.2cm}
\includegraphics[width=0.37\textwidth,angle=90]{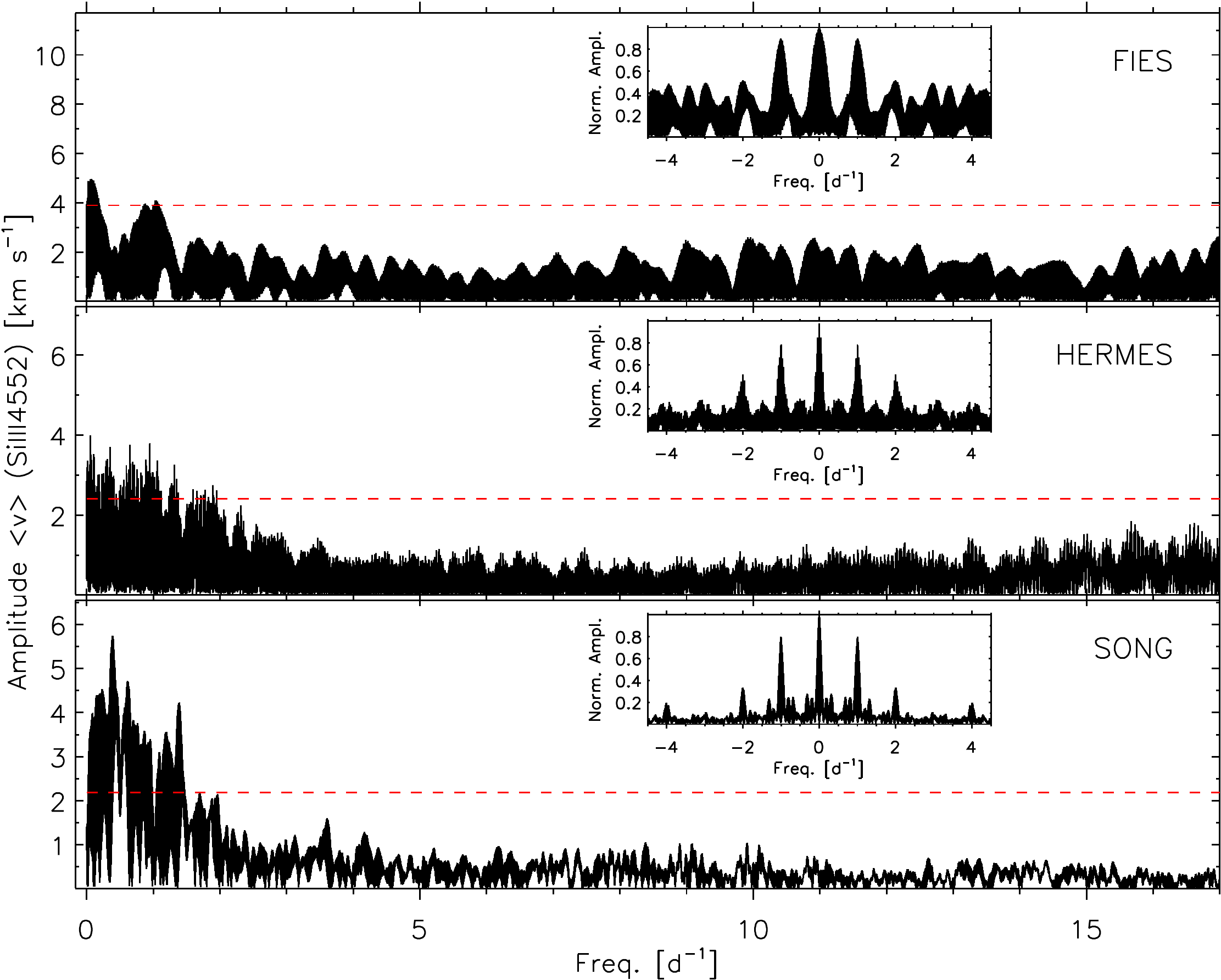}
\caption{Scargle periodograms of the EW (left) and \mom\ (right) of the individual data sets for the \ion{Si}{iii}~$\lambda$4552 line. The corresponding spectral windows are also included as insets. Red horizontal dashed lines indicate four times the average noise level in the range 5\,--\,15 d$^{-1}$.}
\label{Scargle-ew-vrad-4552}
\end{figure*}
\begin{figure*}[!t]
\centering
\includegraphics[width=0.37\textwidth,angle=90]{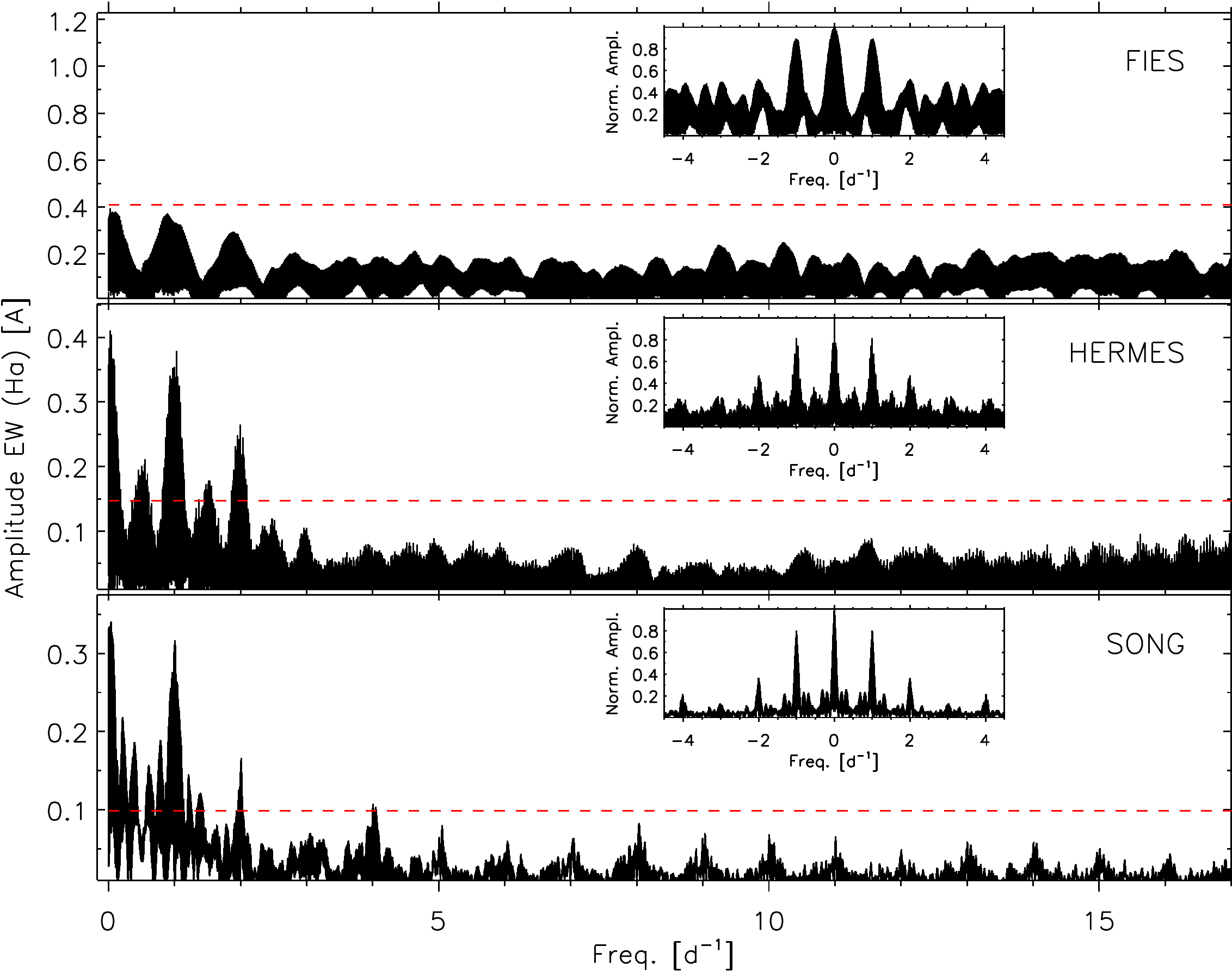}
\hspace*{0.2cm}
\includegraphics[width=0.37\textwidth,angle=90]{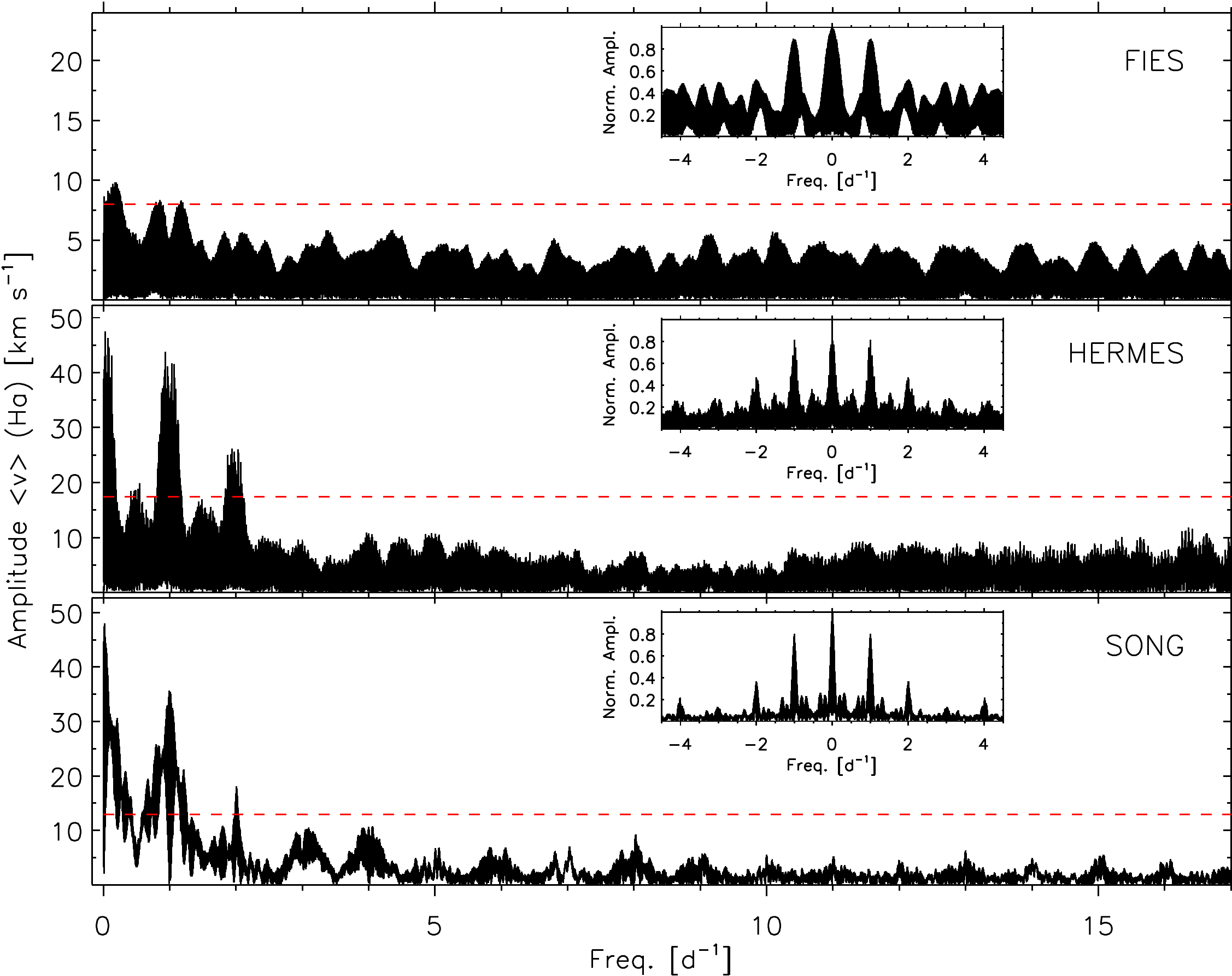}
\caption{Same as Fig.~\ref{Scargle-ew-vrad-4552} for the H$_{\alpha}$ line.}
\label{Scargle-ew-vrad-6563}
\end{figure*}

Figure~\ref{vandew_ex} shows the \ion{Si}{iii}~$\lambda$4552 and
\ion{Si}{iv}~$\lambda$4116 photospheric lines, and three additional lines
(partially) formed in the stellar wind (\ion{He}{i}~$\lambda$5875, H$_{\beta}$
and H$_{\alpha}$), and also includes the equivalent widths. As indicated above,
all lines show clear variability in EW and \mom\ on time scales shorter than the
considered 10 days. The equivalent widths and the centroids of the lines seems
to vary more or less in anti-phase of each other, but not always. Again, we see
that the amplitude of variability increases from the \ion{Si}{iv} line to
H$_{\alpha}$, passing by the \ion{Si}{iii} line, \ion{He}{i}~$\lambda$5875 and
H$_{\beta}$.  A correlation is found between the amplitude of variability of the
lines and how deep they are formed in the stellar photosphere+wind. This is
entirely similar to the case of HD\,188209 \citep{Aerts17}.

\subsubsection{Frequency analysis}\label{freq_analysis}

We computed Scargle periodograms \citep{Scargle82} of the EW and \mom\ of all 13
spectral lines for each of the three FIES, HERMES, and SONG data sets, as well
as for the merged sets. Figures~\ref{Scargle-ew-vrad-4552} and
\ref{Scargle-ew-vrad-6563} illustrate our findings for the individual datasets
for the \ion{Si}{iii}~$\lambda$4552 and H$_{\alpha}$ lines,
respectively. Similar results in terms of global morphology of the Scargle
periodograms and detected frequencies are found for the various lines included
within one of the two groups discussed above (i.e. mostly photospheric or
partially formed in the wind) for which \ion{Si}{iii}~$\lambda$4552 and
H$_{\alpha}$ can be considered as representative, respectively.

The first conclusion that can be extracted from inspection of the individual
Scargle periodograms is that the FIES dataset alone is not enough to provide any
reliable information despite being the one with the longest time-span
($\approx$\,2800 days, see Table~\ref{obslog}). This is due to 
the extremely low duty cycle of these observations. On the contrary, the
time sampling achieved with HERMES and SONG results in much cleaner
periodograms, although still affected by an important 1-day alias feature, a
common phenomenon for single-site data. Interestingly, both lead to a similar
conclusion as the one inferred from the Scargle periodogram associated with the
{\em Hipparcos\/} photometry: the detected variability -- this time in EW and
\mom\ -- is concentrated at low frequencies. In addition, the better
sampling of the hourly time-scales in this case helps to confirm that the
amplitude excess found in the frequency range 6\,--\,14 d$^{-1}$ in the {\em
  Hipparcos\/} periodogram (Fig.~\ref{Scargle-Hp}) is spurious, as was already
suspected from visual inspection of the SONG high cadence observations
(Sect.~\ref{moments-t} and Fig.~\ref{fig1}).

At first glance, the HERMES and SONG periodograms of the EW of the
\ion{Si}{iii}~$\lambda$4552 (left panel in Fig.\,\ref{Scargle-ew-vrad-4552})
do not lead to one and the same dominant frequency. 
The amplitude excess is mainly concentrated below $\sim$0.4
d$^{-1}$. Due to daily aliasing this feature is then repeated in the ranges
0.6\,--\,1.4 d$^{-1}$ and 1.6\,--\,2.4 d$^{-1}$ with decreasing amplitude. A
similar result is found for the EW of all the other investigated lines, as well
as for the first moment of \ion{He}{i}~$\lambda$5875, H$_{\beta}$ and
H$_{\alpha}$ (see Fig.~\ref{Scargle-ew-vrad-6563} for the case of
H$_{\alpha}$). The SONG-\mom\ periodograms of the purely photospheric lines all
reveal a dominant frequency (see, e.g., right panel in
Fig.\,\ref{Scargle-ew-vrad-4552}). Its actual value is slightly different for
the different spectral lines, but is always between 0.36 and 0.39\,d$^{-1}$,
entirely consistent with $f_1$ in the {\em Hipparcos\/} data. This frequency is
also present in the corresponding HERMES-\mom\ periodograms, but not always with
the maximum amplitude. In addition, the amplitude excess at lower frequencies is
somewhat broader in the periodogram obtained from the HERMES-\mom\ dataset
(compared to SONG-\mom, see middle and bottom right panels in
Fig.~\ref{Scargle-ew-vrad-4552}).

\begin{figure*}[!t]
\centering
\includegraphics[width=0.25\textwidth,angle=90]{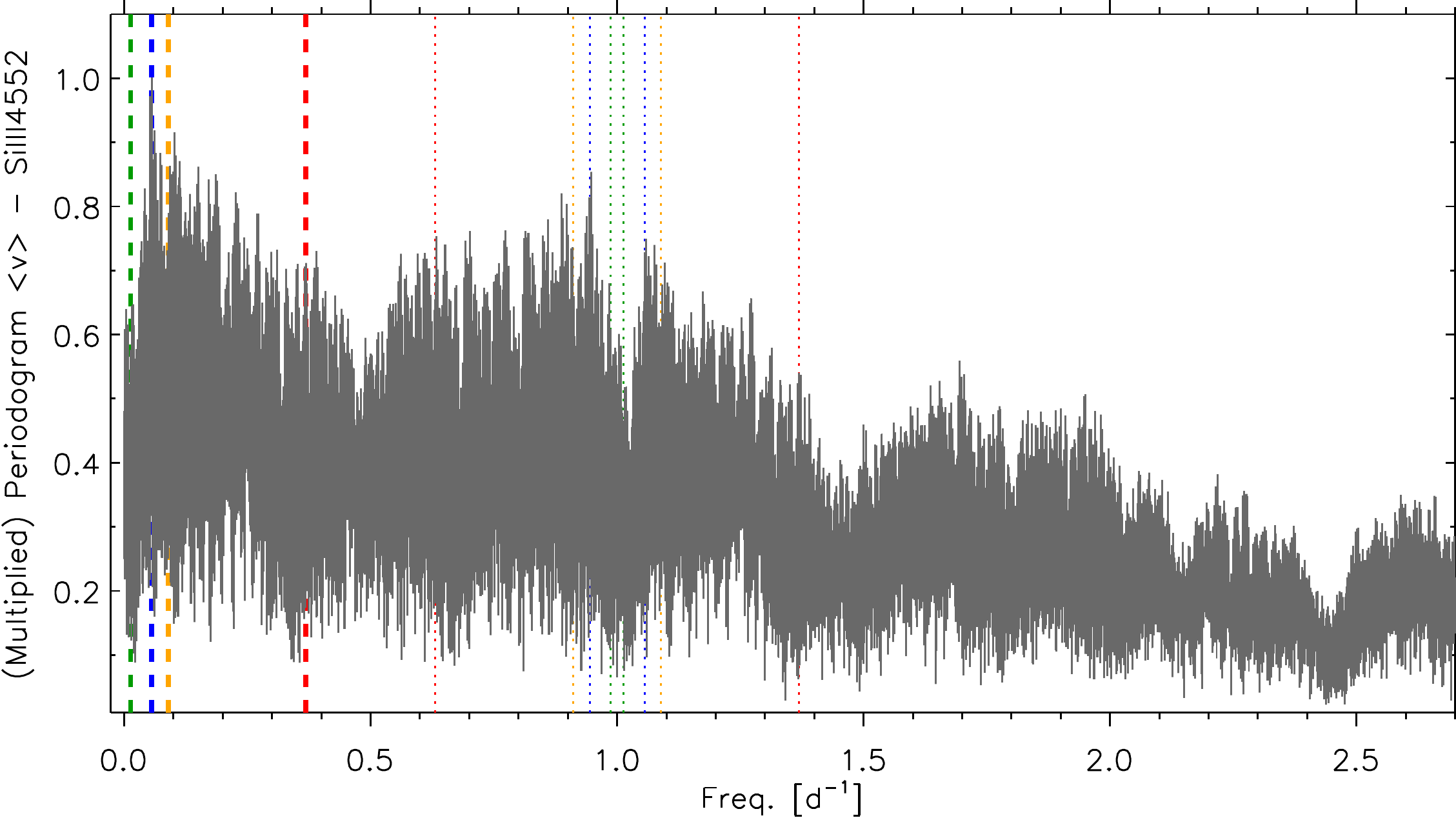}
\hspace*{0.2cm}
\includegraphics[width=0.25\textwidth,angle=90]{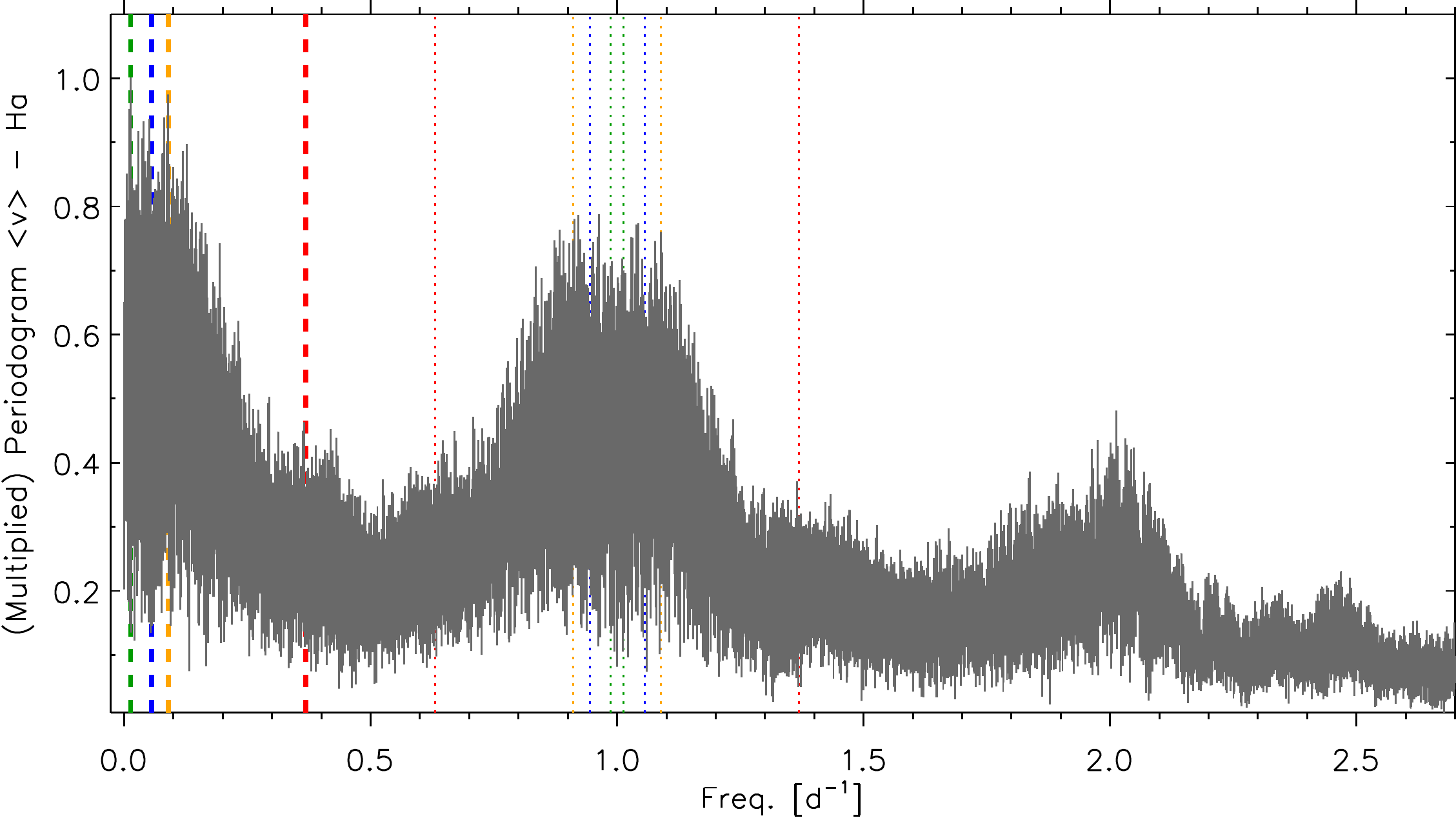}
\includegraphics[width=0.25\textwidth,angle=90]{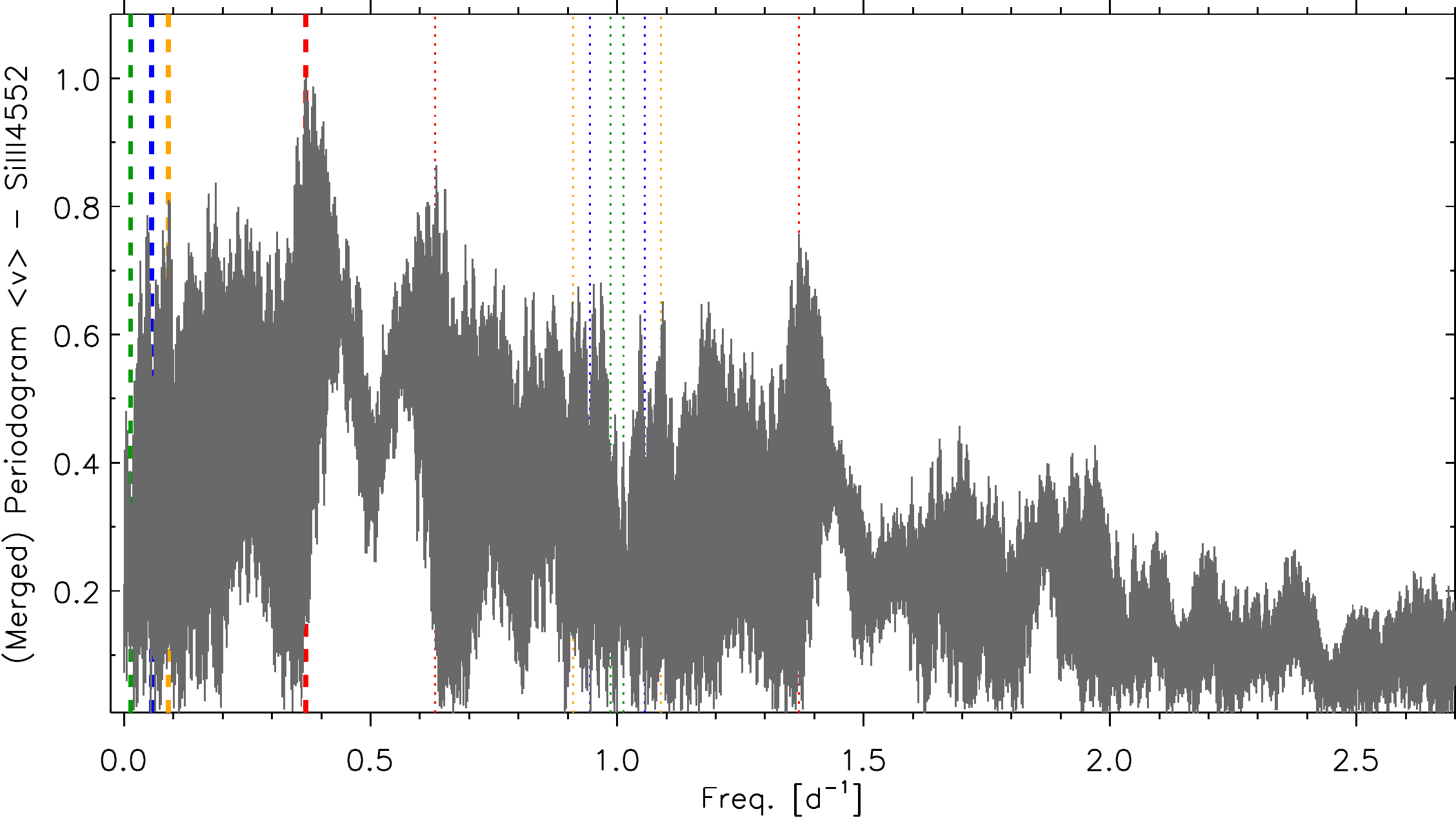}
\hspace*{0.2cm}
\includegraphics[width=0.25\textwidth,angle=90]{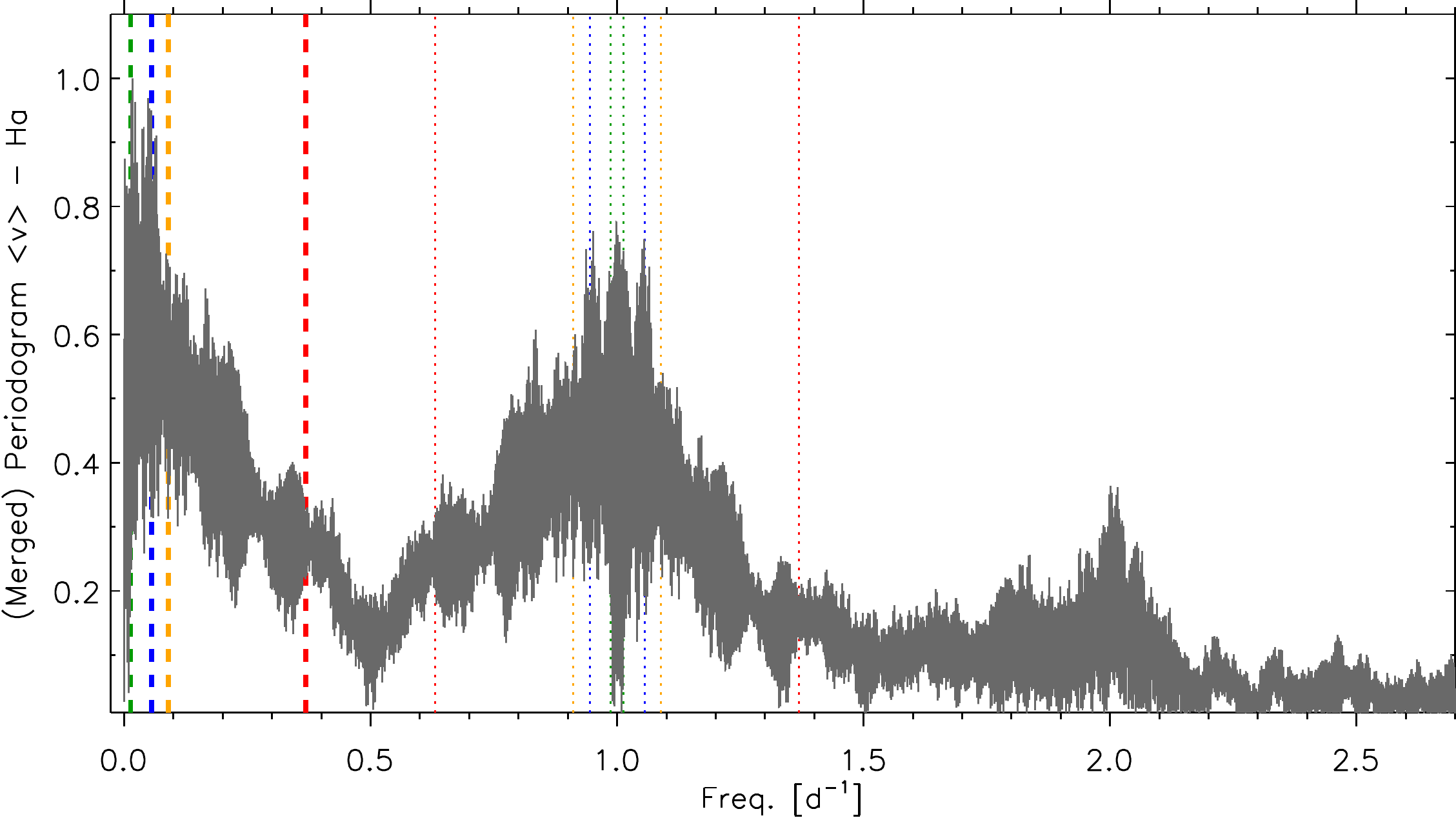}
\caption{Multiplied Scargle periodograms (top panels) of all three individual FIES, HERMES, and SONG data sets after rescaling each one of them to their dominant frequency (see text) compared with the periodogram of the merged data set (bottom panels) for the first moment of the \ion{Si}{iii}~$\lambda$4552 (left) and H$_{\alpha}$ (right) lines. Thick dashed lines indicate some of the dominant frequencies identified in any of the four periodograms. Thin dotted lines show their $\pm$1~day aliases.}
\label{multiplied-Scargle-vrad-4552+Ha}
\end{figure*}

Similarly to \citet{Aerts17}, we also computed multiplied Scargle periodograms by considering the geometric mean per frequency, i.e., we multiplied the periodograms after rescaling each of the individual ones such that the dominant peak receives value 1 and taking the cube root. This method exploits optimally the different structure in the periodograms of FIES and HERMES, where the long-term stability and time base are the asset, and the high cadence of the SONG spectroscopy.  This multiplied periodogram is compared with the one obtained from the merged \mom\ data for the \ion{Si}{iii} line at 4552\AA\ and H$_{\alpha}$ in Fig.\,\ref{multiplied-Scargle-vrad-4552+Ha}. The overall morphology of the four periodograms is consistent and points again to low-frequency amplitude excess (below $\sim$0.45 and $\sim$0.25 d$^{-1}$ for \ion{Si}{iii}~$\lambda$4552 and H$_{\alpha}$, respectively) that is replicated at higher frequencies due to daily aliasing.

%
%
Interestingly, the peak at $\sim$0.37~d$^{-1}$, dominant in the merged periodogram of
\ion{Si}{iii}~$\lambda$4552 -- and marked with a red thick dashed line --, becomes much weaker in
the multiplied periodograms. Further, a new frequency peak at
$\sim$0.05~d$^{-1}$ (blue thick dashed line) becomes dominant in the
latter. This peak is also present in the periodogram of the H$_{\alpha}$ line,
competing in amplitude with  two others at $\sim$0.013 and 0.09 d$^{-1}$ (green
and yellow thick dashed lines, respectively). These four peaks, along with many
other additional ones with lower amplitude -- but clearly above 4 times the
average noise level in the range 5\,--\,15~d$^{-1}$ -- shape the
low-frequency amplitude excess characterizing the variability of  HD\,2905.

\begin{figure}[!t]
\centering
\includegraphics[width=0.45\textwidth,angle=0]{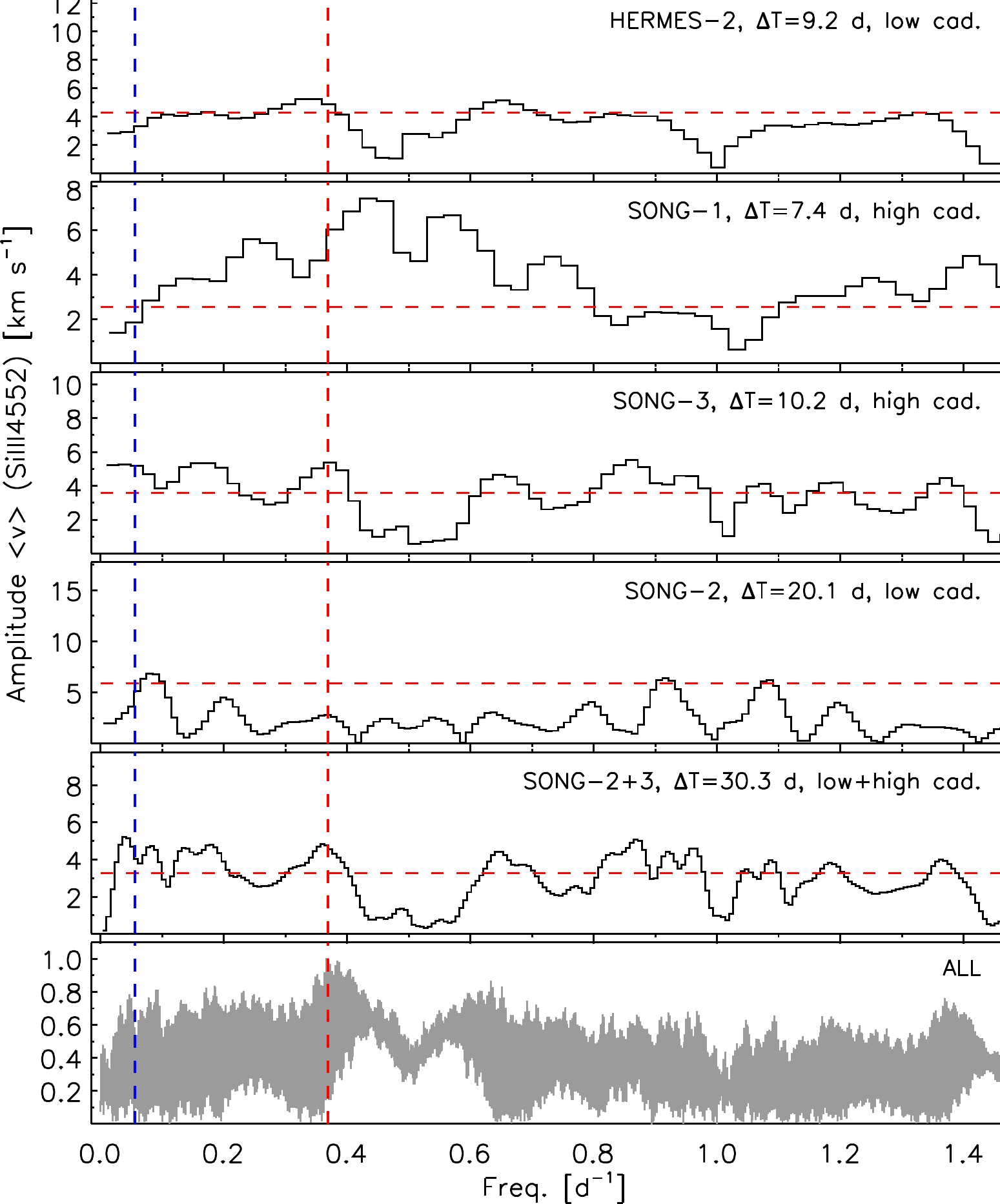}
\caption{Scargle periodograms of the first moment of the \ion{Si}{iii}~$\lambda$4552 line from several of the HERMES and SONG individual runs (see Table~\ref{obslog}) and the complete dataset (bottom panel, see notes in Fig.~\ref{multiplied-Scargle-vrad-4552+Ha}).}
\label{Scargle-vrad-4552-various}
\end{figure}

\begin{figure*}
\centering
\includegraphics[width=0.32\textwidth,angle=-90]{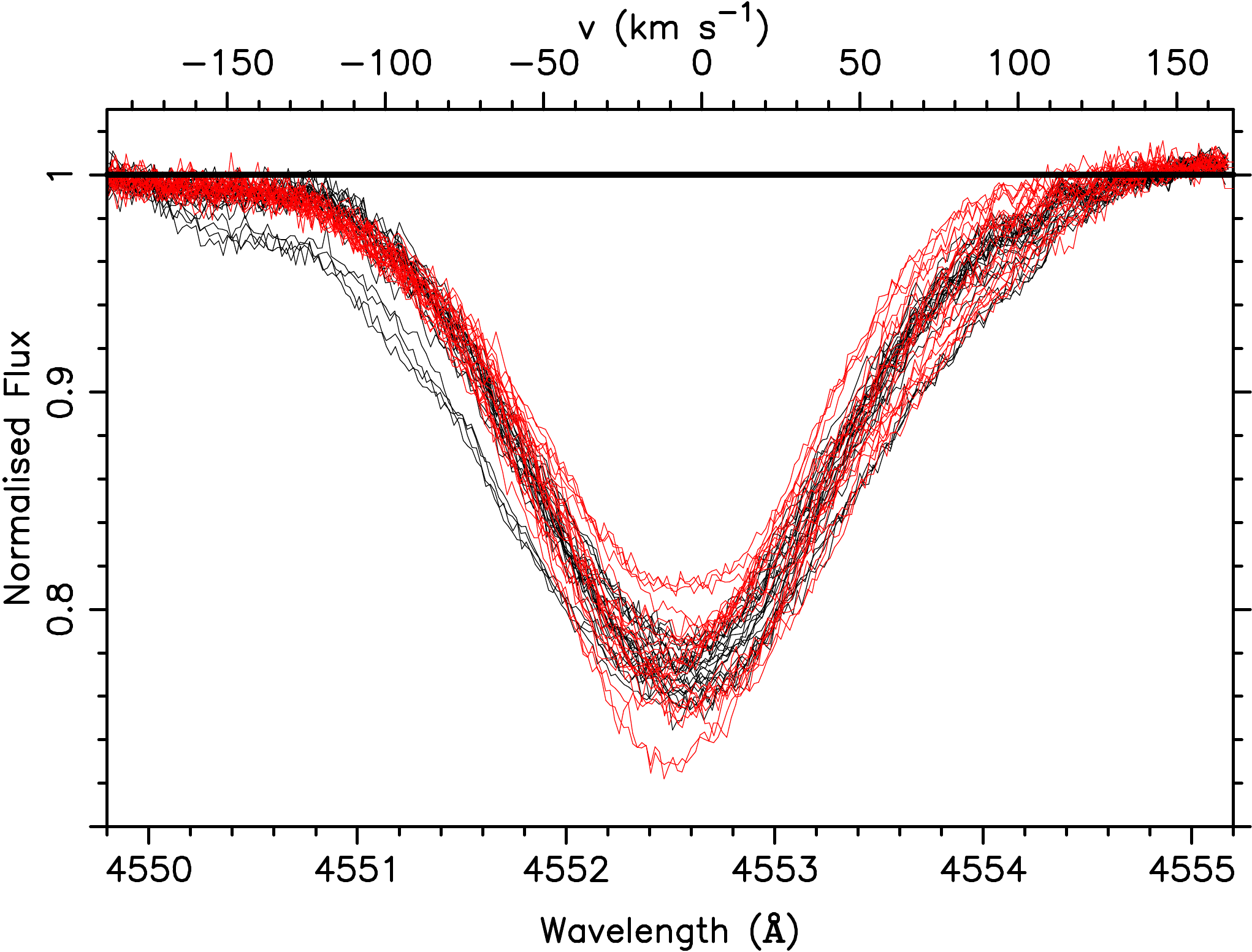}
\includegraphics[width=0.32\textwidth,angle=-90]{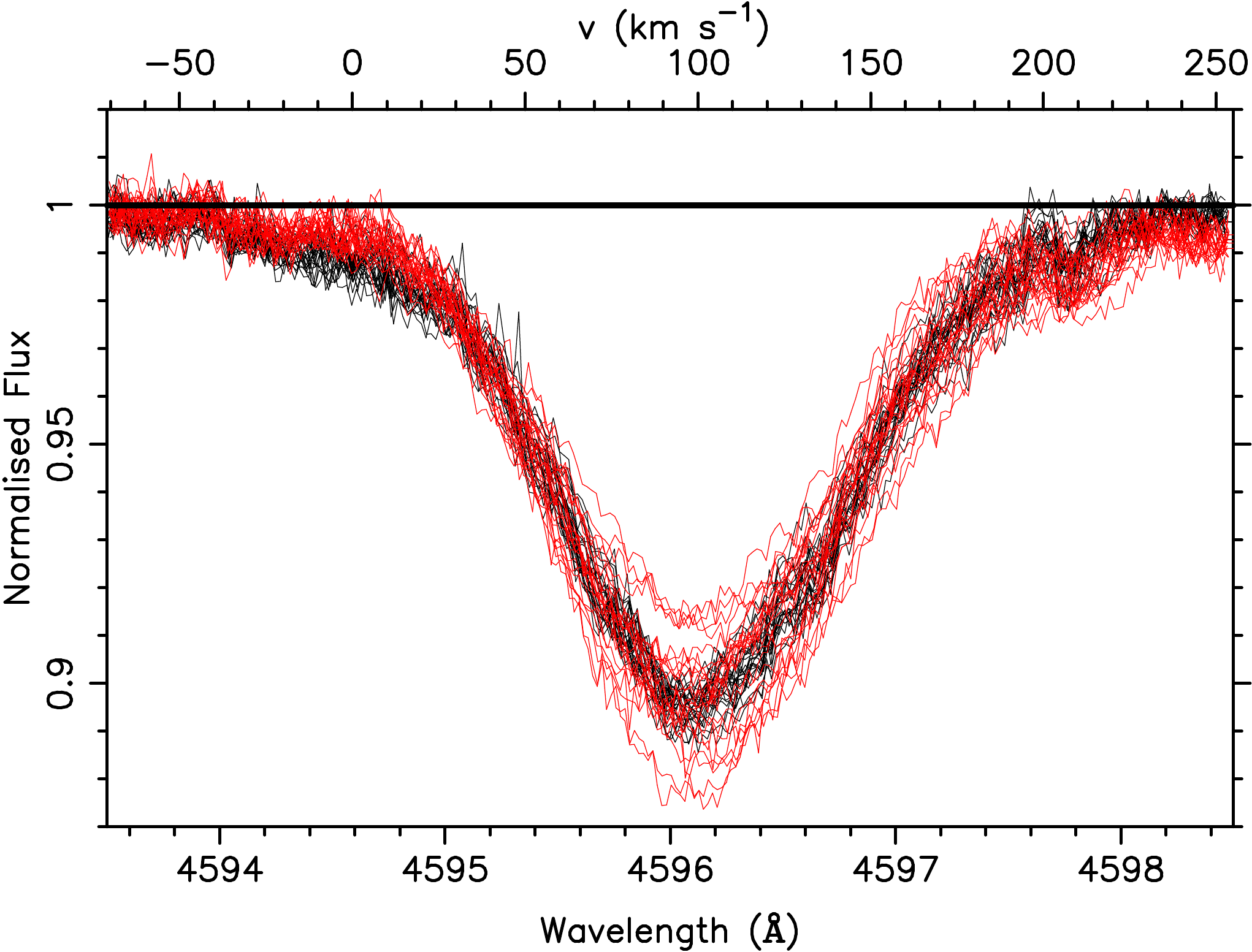}\\[0.1cm]
\includegraphics[width=0.32\textwidth,angle=-90]{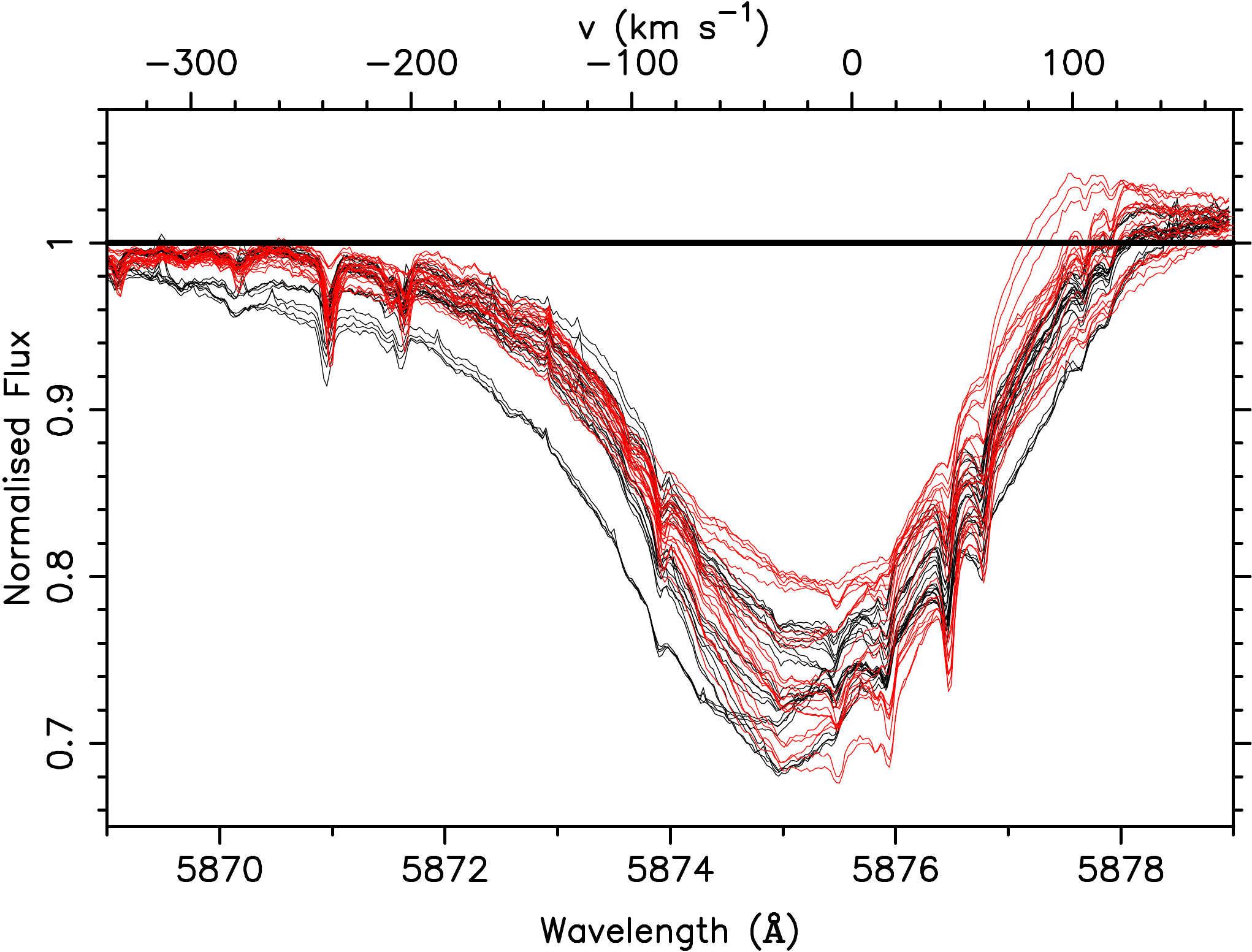}
\includegraphics[width=0.32\textwidth,angle=-90]{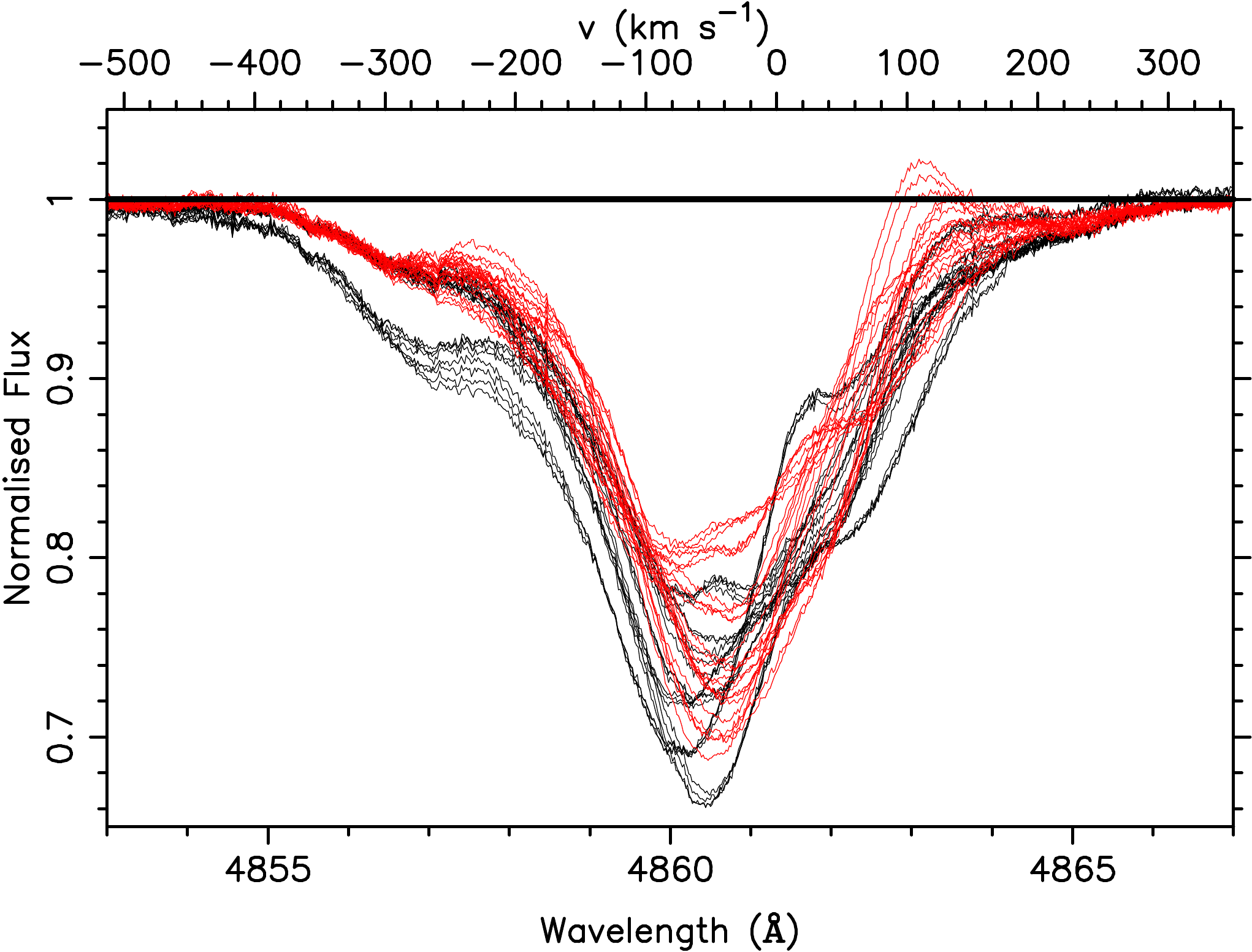}
\caption{Line profiles averaged over 16 individual consecutive measurements of
  the SONG spectroscopy of the first (black) and second (red) intensive
  campaigns. The top two panels are for photospheric lines (left: \ion{Si}{iii},
  right: \ion{O}{ii}) while the bottom two panels are for lines affected by the
  stellar wind (left: \ion{He}{i} with sharp dips caused by telluric lines,
  right: H$\beta$).}
\label{LPV-averaged}
\end{figure*}

\cite{Aerts17} further investigated the nature of variability of the O9~Iab star
HD\,188209 by computing short-time Fourier transformations (STFTs) of the {\it
  Kepler\/} scattered light photometry. This was possible thanks to the
unprecedented space photometric dataset obtained at high cadence for this
star. Such STFTs allow to decide whether we are dealing with multi~mode beating
of stable phase-coherent non-radial pulsation modes \citep[see, e.g., Fig. 5
in][]{Degroote12} and/or a dominant base frequency due to rotational modulation
\citep{Degroote11}, 
or rather variability produced by a stochastic phenomenon. Examples of the
latter can be found in \citet[][Fig.~3]{Tkachenko12}, \citet[][Fig.~6]{Blomme11}
and \citet[][Fig.~5]{Aerts17}. Unfortunately, our spectroscopic dataset is not
suitable for an STFT analysis due to too low cadence for the longest-term data
sets. 

The TESS mission will offer a good opportunity to obtain the appropriate
time-resolved space photometry\footnote{However, only limited to a global
  time-span of $\sim$27~d.} of HD\,2905 to be able to compute meaningful
STFTs. Meanwhile, we present in Fig.~\ref{Scargle-vrad-4552-various} a
comparison of various Scargle periodograms obtained from several of the HERMES
and SONG individual runs (see Table~\ref{obslog}), as well as the complete
dataset. We indicate, for reference, the two dominant peaks found in the merged
and multiplied periodograms for the first moment of the
\ion{Si}{iii}~$\lambda$4552 line (0.3686 and 0.0558 d$^{-1}$,
respectively). While no firm conclusions can be extracted about the temporal
stability of the dominant frequency peaks, as anticipated,
Fig.~\ref{Scargle-vrad-4552-various} illustrates the importance of
assembling an overall spectroscopic dataset that contains high cadence
spectroscopy interlaced in between lower cadence runs with a time base of
several years.

\subsubsection{Line-profile variability}\label{lpv}

\begin{figure*}[!t]
\centering
\includegraphics[width=0.32\textwidth]{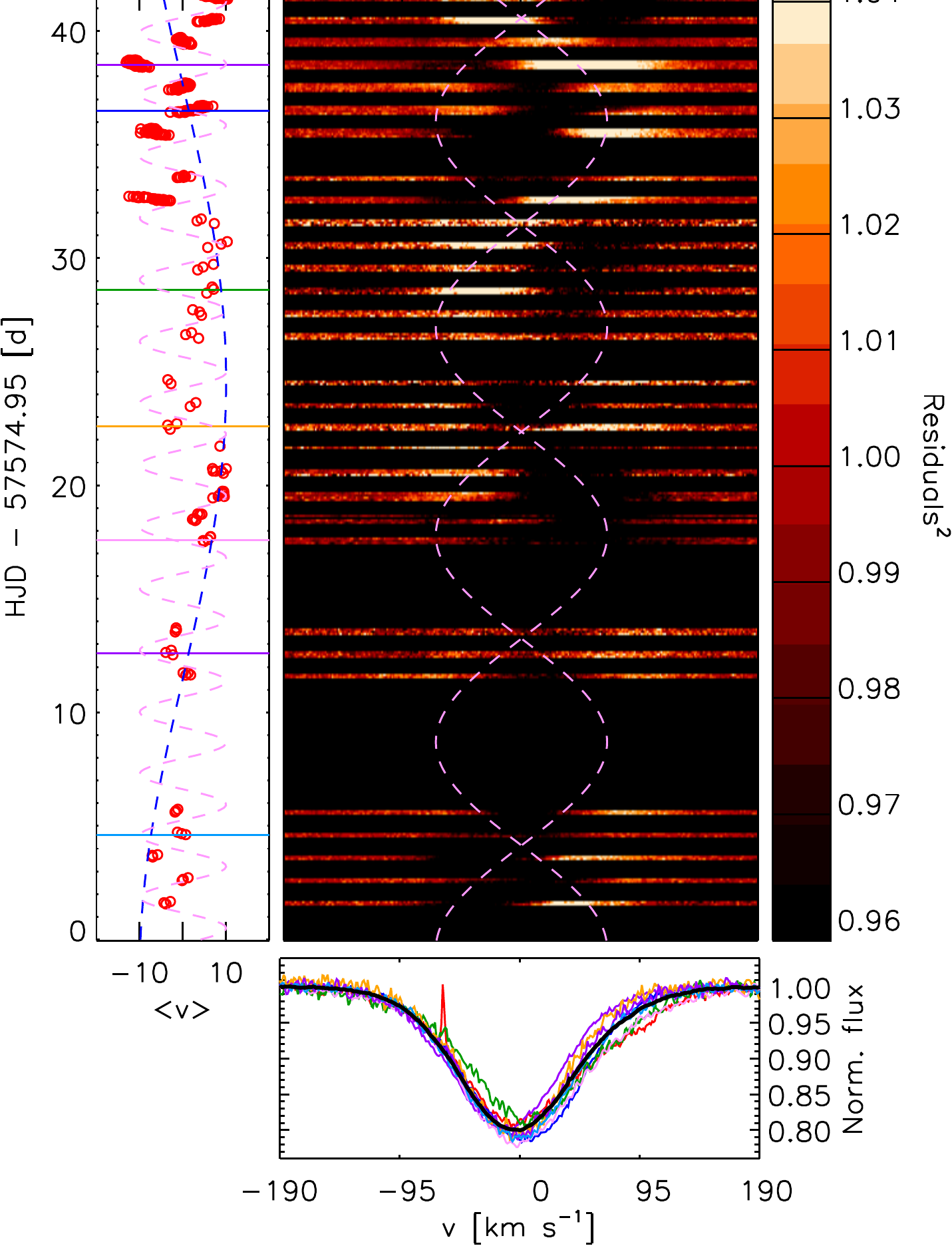}
\hspace*{0.1cm}
\includegraphics[width=0.32\textwidth]{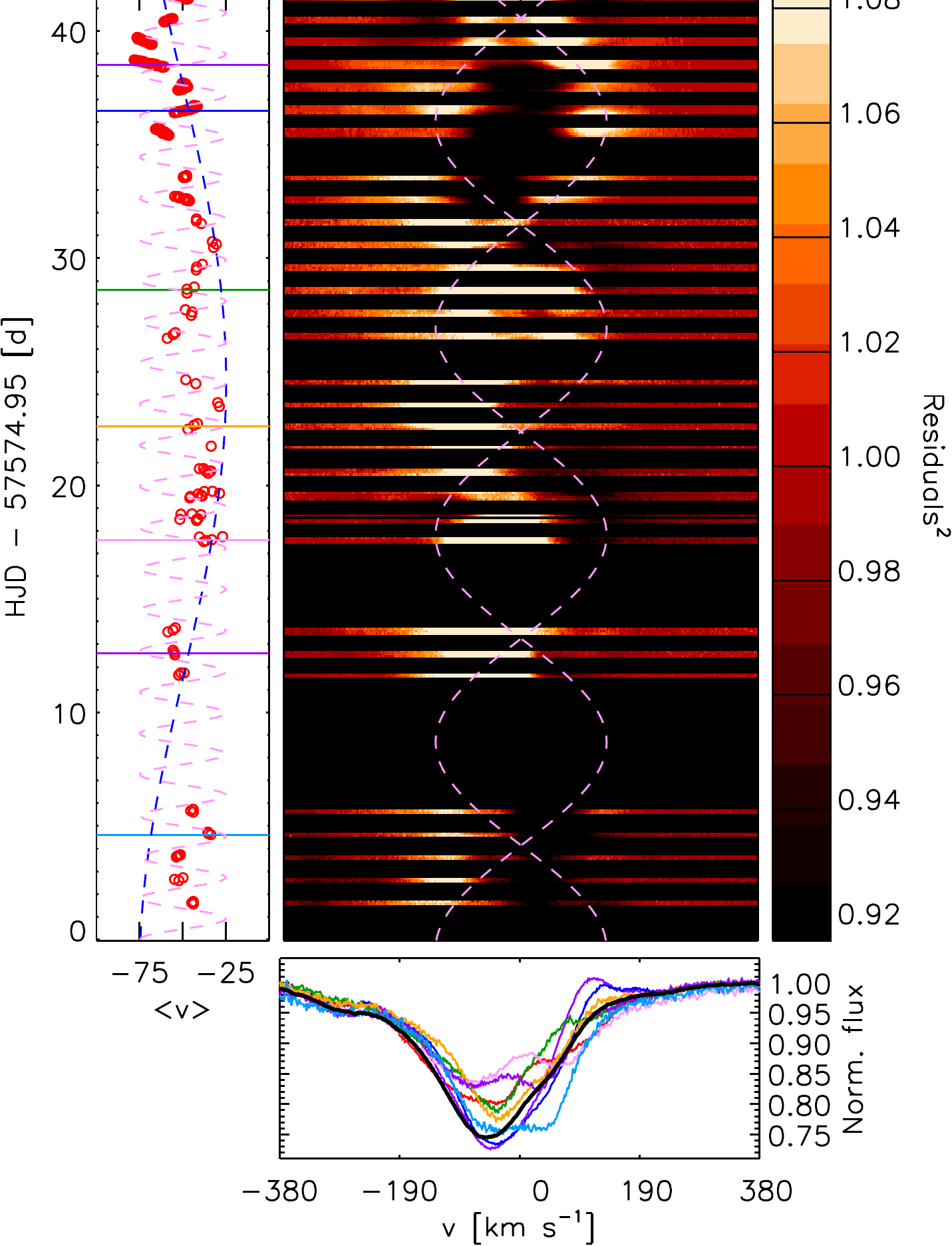}
\hspace*{0.1cm}
\includegraphics[width=0.32\textwidth]{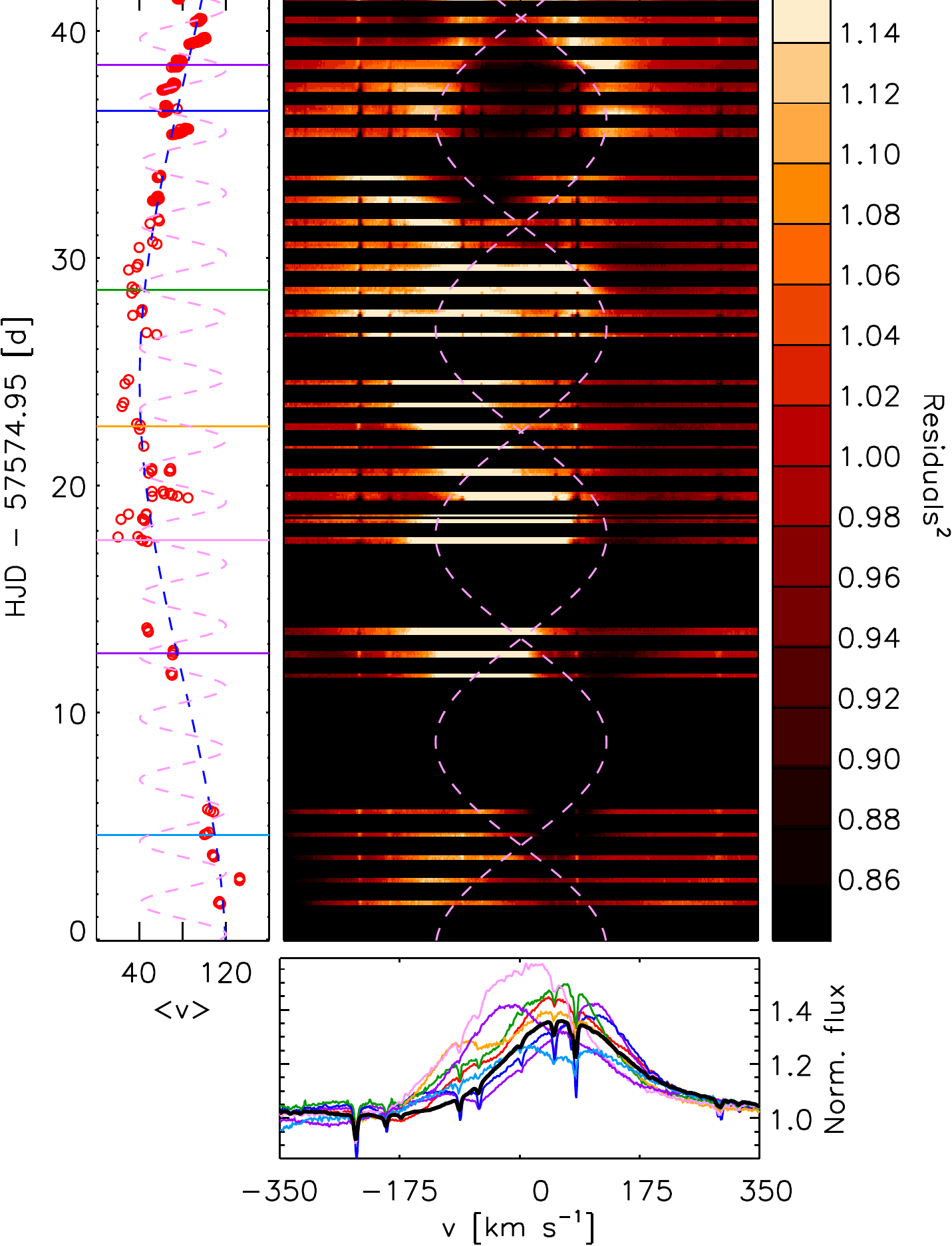}
\caption{Temporal distribution of residuals in $\lambda$-space for three
  representative line-profiles in HD\,2905. (Left) \ion{Si}{iii}~$\lambda$4567 as
  an example of a purely photospheric line. (Right) H$_{\alpha}$ as the most
  extreme case of optical line affected by stellar wind. (Middle) H$_{\beta}$ as
  an intermediate case. We also include for reference purposes three sinusoidal
  curves with associated frequencies 0.055 d$^{-1}$ (pink), 0.019 d$^{-1}$
  (blue) and 0.369 d$^{-1}$ (purple), respectively (see Sect.~\ref{lpv} for
  discussion). Panels to the left shows the temporal variation of the centroid
  of each line-profile (\mom). Bottom panels include several individual profiles
  corresponding to the dates indicated as horizontal lines in the left panels,
  as well as the mean profile obtained from the complete FIES+HERMES+SONG
  spectroscopic dataset.  }
\label{residuals-lpv}
\end{figure*}

In order to further interpret the dominant line-profile variability, which
occurs on time scales of at least a few days, without relying on integrated
quantities such as the equivalent width or centroid velocity, we increased the
S/N ratio of the individual spectral lines by averaging 16 consecutive SONG
spectra. In that way, the S/N is upgraded with a factor 4 and the integration
time is between 3 and 4 hours, which is still only 5 to 7\% of the dominant
periodicity found in the {\em Hipparcos\/} and SONG data. The results of such
averaging are represented for four spectral lines in Fig.\,\ref{LPV-averaged},
where the black/red lines are for the first/second intensive SONG campaigns.  In
this way, line-profile variability is prominently revealed, a typical situation
for pulsations of hot massive stars when studied in spectroscopy of S/N above
250.

All spectral lines vary consistently with each other and lead again to the
conclusion that HD\,2905 has complex multi~periodic line-profile
variability. The variability is dominantly present in the blue spectral line
wing for the first SONG campaign while in the right line wing during the second
SONG run. Despite the presence of telluric lines in the \ion{He}{i} line, we
find the same results for that line and also for H$\beta$, which are both formed
partially in the stellar wind. This is empirical evidence that the line-profile
variability originates in the photosphere but persists in the wind and occurs
with a multiple beating phenomenon that is longer than a week.

Further insight into the characteristics of the spectroscopic variability
present in HD\,2905 can be obtained from the investigation of the residuals in
$\lambda$-space (i.e. waveleght space) of the various diagnostic line profiles. To this aim, we
computed the mean line profile using the complete HERMES+FIES+SONG dataset and
divided each of the individual lines by this mean profile. In the case of the
SONG spectra obtained in high cadence mode, as commented above, we averaged the
resulting residual spectrum within a time-slot of $\sim$2~h (i.e. 16 spectra) to
increase the signal-to-noise ratio. Figure~\ref{residuals-lpv} presents three
cases of lines formed in different regions of the stellar photosphere+wind. The
figure shows the observations gathered in June/July 2016
(HJD\,=\,2457576\,--\,2457617~d), which include the HERMES-4 (4 nights, low
cadence), FIES-3 (3 nights, low cadence), SONG-2 (20 nights, low cadence) and
SONG-3 (10 nights, high cadence) campaigns. When joined together, they can be
considered as the longest observing run with (almost) full coverage of
consecutive nights. Apart from the temporal distribution of residuals in
$\lambda$-space (central panels), we show the centroid velocity curve (\mom,
left panels) and some illustrative line-profiles (bottom panels) for a few
selected dates. As a reference, we overplot in each of the diagrams three
sinusoidal curves corresponding to three different frequencies. The first two
are the dominant peaks found in the merged and multiplied periodograms of the
complete spectroscopic dataset for the \ion{Si}{iii}~$\lambda$4552 line, i.e.,
0.055 and 0.369 d$^{-1}$, respectively (see left panel in
Fig.~\ref{multiplied-Scargle-vrad-4552+Ha}). This second frequency is the
dominant one in the {\em Hipparcos\/} photometry. The third one (0.019 d$^{-1}$)
refers to the dominant frequency standing out from the analysis of the first
moment of the H$_{\alpha}$ line during the $\approx$~40 nights\footnote{Actually, this 
frequency was obtained considering a much larger dataset, also including the HERMES-5 
and HERMES-6 observing runs. In this case, the total time-span of the observations increases 
to 108~d.} considered in Fig.~\ref{residuals-lpv}.

Once more, the diagrams presented in Fig.~\ref{residuals-lpv} show the complex
multi~periodic variability characterizing the line profiles of HD\,2905. Apart
from the almost perfect resemblance of the residuals of the H$_{\alpha}$ and
H$_{\beta}$ lines (despite one is in emission and the other in absorption) the
periodicity associated with the frequency of 0.055 d$^{-1}$ ($\approx$~18~d,
pink curves) seems to be a common feature in the residuals of all analyzed
lines, independently of where they are formed in the outer layers of the
star. In addition, while spectroscopic signatures related to the other two
quoted frequencies seem to be present in the \mom\,--\,curves of the three
lines, 0.369 d$^{-1}$ ($\approx$~2.7~d, purple curves) dominates the temporal
behavior of the photospheric lines and the photometry, while the large scale
wind variability is mainly modulated by the frequency 0.019 d$^{-1}$
($\approx$50~d, blue curves).

\section{Discussion}\label{section5}

Single-site\footnote{Although we have use three different telescopes, the data
  can be considered as a single site set since all three are located at the
  Canary Islands' observatories.}  spectroscopy of the B1Ia star HD\,2905
gathered during $\sim$7 years in campaigns with a duration of 4\,--\,20 nights
revealed empirical evidence of photospheric and wind variability with
periodicities of a few to several days. One dominant frequency of
$\sim 0.37\,$d$^{-1}$ was found in both the {\it Hipparcos\/} space photometry
and in the photospheric spectral lines.  Aside from this frequency, HD\,2905
reveals multi-periodic spectroscopic variability. The overall peak-to-peak
amplitude in the zero and first moments of the photospheric lines reach up to
15\% and 30~\kms, respectively. The variability increases in amplitude for lines
formed deep in the photosphere to those formed in the stellar wind. This
property was also found for the O9.5Iab star HD\,188209 \citep{Aerts17}. 

The first moments of the \ion{Si}{iii}$\lambda$4552 and H$\alpha$ lines of
HD\,2905 reveal variations with frequencies below $\sim$0.4 and $\sim$0.1
d$^{-1}$, respectively. In addition, several frequency peaks located in the
range $\sim$0.02\,--\,0.4~d$^{-1}$ occur on top of a global amplitude excess in
the various periodograms.  The frequencies below $\sim$0.05 d$^{-1}$ have a
large amplitude in the periodograms of the H$\alpha$ line.  If we assume
$v_{\rm eq}$\,=\,60~\kms\ and $R$\,=\,40~$R_{\odot}$ (see
Table~\ref{parameters_log}), we obtain P$_{\rm rot}$=34~d (or, equivalently,
f$_{\rm rot}$\,=\,0.029 d$^{-1}$). Taking into account that we are actually
measuring $v_{\rm eq}$~sin~$i$, but also that there is some empirical evidence
indicating that our current techniques to derive projected rotational velocities
in B supergiants lead to upper limits of this quantity \citep[][]{Simon14}, we
assess that any rotational variability should occur in the frequency domain
below $\approx$0.1 d$^{-1}$. Therefore, the two dominant frequency peaks located
at $\approx$0.019 and $\approx$0.055~d$^{-1}$ (see Sect.~\ref{freq_analysis} and
Figures~\ref{multiplied-Scargle-vrad-4552+Ha} and \ref{residuals-lpv}) could be
related to the rotation of the star. The longer term variability among these two
($\approx$50~d) is only detected in the H$\alpha$ line, which must be formed in
the stellar wind as it permanently reveals strong emission.  This result,
combined with the time dependent behavior of its shape, points to a rotationally
modulated non-spherical stellar wind as the main cause of this aspect of the
detected variability.  Line-profile variability associated with the second
low-frequency peak at $\approx$0.055 d$^{-1}$ ($\approx$20~d) is detected in the
residuals of H$\beta$ and the \ion{Si}{iii}$\lambda$4552 line (see
Fig.~\ref{residuals-lpv}).  Indeed, this frequency dominates in the multiplied
Scargle periodogram of this latter line (see left panel in
Fig.~\ref{multiplied-Scargle-vrad-4552+Ha}). Therefore, rotational modulation
may also explain part of the variability detected in lines formed in the
photosphere, while these lines are dominated by shorter-term variability at
$\sim 0.37$d$^{-1}$ and other candidate frequencies above 0.1\,d$^{-1}$.

It remains to be studied whether the photospheric variability of HD\,2905 found
in the range $\approx$~0.1\,--\,0.4 d$^{-1}$ (see
Fig.~\ref{multiplied-Scargle-vrad-4552+Ha}) is of stochastic nature or
not. Aside from this, the quasi-stable frequency $\sim 0.37\,$d$^{-1}$ seems to
point towards a phenomenon of coherent variability as it persists over years in
both the spectroscopy and the space photometry. A combination of stable
heat-driven gravity modes \citep{Godart17} with convectively-driven stochastic
gravity waves excited in the stellar interior -- either in the convective core
\citep{Aerts15} or in the convection zone in the stellar envelope due to the
iron opacity bump \citep{Grassitelli15} -- seems to offer the best explanation 
for the complex variability detected in the spectroscopy of HD\,2905.

As in the study of the late O supergiant HD\,188209 by \cite{Aerts17},
uninterrupted high-cadence space photometry combined with long-term
high-precision spectroscopy is the best strategy to tackle the full
interpretation of the complex variability of B supergiants. In this work, we
managed to distinguish between the weeks-long wind variability and the days-long
photospheric variability. Moreover, we found the latter to propagate into the lower part
of the stellar wind.  In order to understand the full cause of the photospheric
variability, 3D simulations of gravity waves on the one hand and of convection
in the sub-surface layers on the other hand are needed. The frequency spectra of
the photospheric motions predicted by these two distinct phenomena can then be
compared with the measured frequency spectra of a large sample of blue
supergiants to come to a solid physical interpretation of the overall
variability.

\begin{acknowledgements}
  Based on observations made with the Nordic Optical Telescope, operated by
  NOTSA, and the Mercator Telescope, operated by the Flemish Community, both at
  the Observatorio del Roque de los Muchachos (La Palma, Spain) of the Instituto
  de Astrof\'isica de Canarias. Based on observations made with the Hertzsprung
  SONG telescope operated on the Spanish Observatorio del Teide on the island of
  Tenerife by the Aarhus and Copenhagen Universities and by the Instituto de
  Astrof\'isica de Canarias.  This work has been funded by the Spanish Ministry
  of Economy and Competitiveness (MINECO) under the grants AYA2010-21697-C05-04,
  AYA2012-39364-C02-01, and Severo Ochoa SEV-2011-0187, by the European Research
  Council (ERC) under the European Union's Horizon 2020 research and innovation
  programme (grant agreement N$^\circ$670519: MAMSIE), and by the National
  Science Foundation of the United States under Grant N$^\circ$NSF
  PHY11--25915. We thank Jens Jessen-Hansen, Ren\'e Tronsgaard Rasmussen and
  Rasmus Handberg for their work on the SONG data-reduction pipeline and the
  SONG archive system. SS-D kindly acknowledges the staff at the NOT and
  Mercator telescopes for their professional competence and always useful help
  during the observing nights. We thank the referee for the time spent in carefully 
  reading the paper and her/his suggestions for improvement of the text.
  \end{acknowledgements}

\end{document}